\documentclass{aa}
\usepackage{graphicx,amsmath,txfonts}

\begin{document}
\newcommand{\mearth}{M_\oplus}
\title{Migration and gas accretion scenarios for the Kepler 16, 34 and 35 circumbinary planets}
\author{ A.Pierens \inst{1,2} \and R.P. Nelson \inst{3} }
\institute{ Universit\'e de Bordeaux, Observatoire Aquitain des Sciences de l'Univers,
    BP89 33271 Floirac Cedex, France \label{inst1} \and
   Laboratoire d'Astrophysique de Bordeaux,
    BP89 33271 Floirac Cedex, France \label{inst2} \\
     \email{arnaud.pierens@obs.u-bordeaux1.fr}
 \and Astronomy Unit, Queen Mary University of London, Mile End Road, London, E1 4NS, UK
 \label{inst3} }
\abstract
{Several circumbinary planets have been detected by the Kepler mission. 
Recent work has emphasized the difficulty of forming
these planets at their observed locations due to perturbations by the binary. It has been suggested that these planets formed further out in their discs
in more quiescent environments and migrated in to locations where they are observed.}
{We examine the orbital evolution of planets embedded in circumbinary disc models for the three systems
Kepler-16, Kepler-34 and Kepler-35. The aims are: to explore the plausibility of a formation scenario in which cores form at large
distances from the binaries and undergo inward migration and gas accretion as the gas disc disperses; to 
determine which sets of disc parameters lead to planets whose final orbits provide reasonable fits to the observed systems.}
{Using a grid--based  hydrodynamics code we performed simulations of a close binary system interacting with 
circumbinary discs with differing aspect ratios, $h$, and viscous stress parameters $\alpha$.  
Once the binary+disc system reaches quasi-equilibrium we embed a planet in the disc and examine its 
evolution under the action of binary and disc forces. We consider fully-formed planets with masses equal to those inferred from Kepler data,
and low-mass cores that migrate and accrete gas while the gas disc is being dispersed.}
{A typical outcome for all systems is stalling of inward migration as the planet enters the tidally-truncated inner cavity
formed by the binary system. The circumbinary disc becomes eccentric through interaction with the binary, and the disc
eccentricity forces the planet into a noncircular orbit. For each of the Kepler-16b, Kepler-34b and Kepler-35b
systems we obtain planets whose parameters agree reasonably well with the observational data, but none of our
simulations are able to produce highly accurate fits for all orbital parameters.} 
{The final orbital configuration of a circumbinary planet is determined by a delicate interplay between
the detailed stucture of the circumbinary disc and the orbital parameters of the planet as it migrates
into the inner disc cavity. Simplified simulations such as those presented here provide support for a formation
scenario in which a core forms, migrates inward and accretes gas, but accurate fitting of the observed Kepler
systems is likely to require disc models that are significantly more sophisticated in terms of their input physics.}
\maketitle

\section{Introduction}
Among the thousands of planet candidates discovered by Kepler, at the time of writing  
6 of them are circumbinary planets which orbit around both components of a close binary system
on a so-called P-type orbit. Kepler-16b was the first circumbinary planet discovered 
with Kepler (Doyle et al. 2011). It is a Saturn-mass planet with near-circular 
orbit and semi-major axis $a_p \sim 0.7$ AU (measured from the centre of mass of the binary). 
The binary harbouring Kepler-16b has semi-major axis $a_b\sim 0.22$ AU,  eccentricity $e_b\sim 0.16$ 
and mass ratio $q_b=M_B/M_A\sim 0.29$, where $M_A$ and $M_B$ are the masses of the primary and secondary stars respectively. 
Interestingly, Kepler-16b is found to orbit just outside the long-term stability limit  
located at $\sim 0.64$ AU from the binary (Holman \& Wiegert 1999). Two additional transiting circumbinary planets 
have been reported by Welsh et al. (2012) around Kepler-34 and Kepler-35: binary systems with stellar binary periods of $28$ and $21$ days, respectively. 
Kepler-34b has orbital period $289$ days and mass ratio $q=m_p/M_\star\sim 1.1 \times 10^{-4}$, where $M_\star=M_A+M_B$ and $m_p$ is the planet mass. 
Kepler-35b has orbital period $131$ days and mass ratio $q\sim 7.5\times 10^{-4}$. The binary and planet parameters for 
these three systems are summarized in Table \ref{table1}. More recent detections include the Neptune-sized circumbinary 
planet Kepler-38b (Orosz et al. 2012a),  a  system of 2 planets around Kepler-47 (Orosz et al. 2012b), and a 
transiting circumbinary planet around KIC 4862625 (Schwamb et al. 2013).

These circumbinary planets are all found to have highly aligned orbit planes with respect to
the central binaries. Although transit photometry is a detection technique biased toward finding coplanar orbits 
(Winn et al. 2011), this nonetheless suggests that these planets formed within a coplanar circumbinary disc. For Kepler-16b,  
this is further supported by the fact that the spin of the primary is aligned with the angular momentum of the binary+planet system 
(Winn et al. 2011). Moreover, this is also consistent with detections of circumbinary discs around spectroscopic binaries like DQ Tau, 
AK Sco and GW Ori. In GG Tau, the circumbinary disc has been directly imaged and has revealed the presence of an inner 
cavity due to the tidal torques exerted by the central binary (Dutrey et al. 1994). 

Early studies of the dynamics of planetesimals embedded in circumbinary discs adopted
simple axisymmetric disc models, and showed that eccentric orbits induced by
the tidal influence of the binary companions would  experience pericentre aligment
due to gas drag forces (Scholl et al. 2007). In principle this can reduce the collision
velocities between planetesimals, but pericentre alignment is size-dependent leading
to the expectation that collisions within a swarm consisting of a range of planetesimal
sizes may be erosive (Scholl et al. 2007). Moreover, non-axisymmetric structures in the disc such
as spiral density waves or global eccentric modes may also lead to large impact velocities,
especially if the gravitational field of the disc is included in the evolution of planetesimal
orbits (Marzari et al. 2008; Kley \& Nelson 2010).

Regarding the circumbinary planets discovered with Kepler, Paardekooper et al. (2012) recently performed N-body 
simulations of planetesimal accretion in these systems, including planetesimal formation and dust accretion. They showed 
that the large eccentricities reached by the planetesimals make it difficult to form these planets in-situ. A similar 
conclusion was drawn by Meschiari (2012) who found that for Kepler-16b, planetesimal accretion might be inhibited 
interior to $\sim 4$ AU. Both studies indicate that these planets formed further out in the disc, 
in a more accretion-friendly environment, and migrated inward to their present location through Type I (Ward et al 1997, Tanaka et al. 2002) 
and/or Type II migration (e.g. Lin \& Papaloizou 1986; Nelson et al. 2000). 

The evolution of Earth-mass bodies embedded in a circumbinary disc and undergoing Type I migration was examined by Pierens \& Nelson (2007). 
They found that the inward drift of a protoplanet can be stopped near the edge of the cavity formed by the binary due to the
strong positive corotation torque there counterbalancing the Lindblad torque (Masset et al. 2006).
The orbital evolution of Jovian mass planets embedded in circumbinary discs was studied by Nelson (2003) and subsequently by 
Pierens \& Nelson (2008).  These authors found that the typical evolution of Saturn-mass planets 
involves inward migration until the planet eccentricity 
becomes high enough to induce a torque reversal and the subsequent outward migration of the planet. This effect makes  Saturn-mass planets 
evolve on a stable orbit, at a safe distance  from the central binary.   
Giant planets 
with masses of $m_p\ge 1$ $M_J$, however,  generally undergo close encounters with the binary and are subsequently completely ejected from the system. 
We notice that the values for the inferred masses of the circumbinary planets discovered with Kepler are surprisingly consistent with these results.

In this paper, we present hydrodynamic simulations of the orbital evolution of planets embedded in circumbinary disc models
designed to mimic the protoplanetary discs that existed in the Kepler-16, Kepler-34 and Kepler-35 systems during the epoch of planet formation.
The models we adopt are by their nature very simple: flat 2D locally isothermal discs with constant $H/R$ values in which
the accretion stress is modelled using the standard alpha prescription (Shakura \& Sunyaev 1973). The aim of this work is to
examine the general plausibility of the ideas presented in our previous papers that the orbital configuration of circumbinary 
planets is determined through their interaction with the circumbinary disc, which develops an inner cavity and becomes eccentric
through tidal interaction with the central binary. In principle, with sophisticated disc modelling that includes processes such as
magnetohydrodynamic turbulence, time dependent radiative heating of the disc by the stellar components, self-gravity, heat transport
and a realistic equation of state, one could attempt to fit the Kepler observations of circumbinary planets and thereby constrain
the nature of the circumbinary discs in which the planets formed and migrated. Such a project is beyond current capabilities
so we adopt the less ambitious aim of obtaining decent agreement between our simplified simulations and the observed Kepler
circumbinary planets to support the plausibility of the general core formation, migration and gas accretion
model for these planets.

We consider two basic scenarios: (i) formation of the planet at large distance from the central
binary followed by migration to a final orbit close to the central binary without further mass growth;  (ii) formation of a 20 M$_{\oplus}$ core 
at large distance that migrates inward through type I migration and then accretes gas once migration is halted, and before disc dispersal determines its 
final mass and orbit. We examine the outcome of simulations as a function of the adopted disc parameters 
($h \equiv H/R$; viscous $\alpha$ parameter; disc-dispersal time scale). 
We find that different disc models are needed to produce decent approximations to the observed parameters of each of the
Kepler-16, Kepler-34 and Kepler-35 systems. For example, a circumbinary disc with $h=0.05$ and $\alpha=10^{-2}$ reproduces 
Kepler-34b reasonably well, whereas $h=0.03$ and $\alpha=10^{-3}$ provide decent agreement with the Kepler-35b data.
For Kepler-16b, a close fit to the observed parameters is obtained using a disc model with $h=0.05$ and $\alpha=10^{-4}$,
although all our simulations have difficulty in reproducing the very low eccentricity reported for this system.
For each system we find that the planet migrates into the tidally truncated disc cavity created by the
binary before halting, as anticipated by Nelson (2003) and Pierens \& Nelson (2007; 2008). In the case of Kepler-16b, 
halting of migration occurs when the amount of local gas in the cavity becomes too small to make the planet migrate 
further in; whereas for Kepler-34b, Kepler-35b, this arises once the planet eccentricity becomes high enough to 
induce a torque reversal.\\

This paper is organized as follows. In Sect. 2, we present the hydrodynamic model. In Sect. 3, we discuss results 
concerning binary-disc interactions. Results of simulations with fully-formed planets are presented in 
Sect. 4, and in Sect. 5  we examine the scenario in which a low-mass embryo grows by 
accreting gas from the disc once it has undergone substantial migration. Finally we discuss our results and
draw our conclusions in Sect. 6.

\section{The hydrodynamical model}
\subsection{Numerical method}
We adopt a 2D disc model in which all physical quantities are vertically averaged 
and work in polar coordinates $(R,\phi)$ with the origin located at the centre of mass 
of the binary.  The equations governing disc evolution and the binary plus planet system 
can be found in Pierens \& Nelson (2007). The simulations presented here were performed 
using the GENESIS numerical code which solves the equations for the disc using an advection 
scheme based on the monotonic transport algorithm (Van Leer 1977). This code has been employed 
in numerous previous studies of planet formation in circumbinary discs (Pierens \& Nelson 2007; 2008a; 2008b). 

Evolution of the planet and binary orbits is computed using a fifth-order Runge-Kutta integrator (Press et al. 1992). 
For moderate values of disc viscosity, exchange of angular momentum with the disc can lead to significant 
modification of the binary orbital elements, and it is generally expected that shrinking of the binary separation 
plus growth of the binary eccentricity will result from the resonant interaction between the disc and binary 
(Pierens \& Nelson 2007). Here, since we want to reproduce as close as possible the observed properties of the Kepler 
circumbinary planets, the force exerted by the disc onto the binary is not included in the evolution equations, so that 
the binary orbital elements remain fixed in the course of the simulations with values corresponding to the observed ones. 

We adopt computational units such that 
the total mass of the binary $M_\star=1$, the gravitational constant $G=1$ and the radius $R=1$ in the 
computational domain corresponds to $0.5$ AU.  When presenting simulation results we report time in units of 
the binary orbital period $P=2\pi\sqrt{GM_\star/a_b^3}$.

 The simulations employ $N_R=432$ radial grid cells uniformly distributed between $R_{in}=1.5\;a_b$, where $a_b$ is the 
binary semi-major axis, and $R_{out}=10$ (5 AU), and $N_\phi=512$ azimuthal grid cells.

\subsection{Initial conditions}
\label{sec:init}
The initial disc surface density profile is chosen to be $\Sigma(R)=f_{gap}\Sigma_0R^{-3/2}$ where $\Sigma_0$ 
was defined such that the disc contains $2 \%$ of the total mass of the binary within $30$ AU and 
$f_{gap}$ is a gap function which is employed to model the inner cavity that is expected to be formed 
as a result of binary-disc interactions. Following Gunther et al. (2004), we set:
$$
f_{gap}=\left(1+\exp\left[-\frac{R-R_{gap}}{0.1R_{gap}}\right]\right)^{-1}
$$ 
where $R_{gap}=2.5\;a_b$ is the estimated gap size (Artymowicz \& Lubow 1994).
We adopt a locally isothermal equation of state  (this assumption is discussed and justified 
in Sect. \ref{sec:valid}) with temperature profile given by 
$T=T_0R^{-\beta}$ where $\beta=1$ and $T_0$ is the temperature at $R=1$. This corresponds 
to a disc with constant aspect ratio $h$ for which we consider values of 
$h=0.03$, 0.05 and 0.07. For a fiducial run with $h=0.05$, the Toomre parameter $Q=c_s \kappa/\pi G \Sigma$, 
where $\kappa$ is the epicyclic frequency and $c_s$ the sound speed, is estimated to be $Q\sim 100$ at the disc inner edge 
and $Q\sim 25$ at the disc outer edge, which implies that we can safely ignore self-gravity in this work. 
Neglecting effects resulting from self-gravity is further justified by the recent results of 
Marzari et al. (2013) who found that for similar values of the Toomre parameter, including 
self-gravity does not significantly affect the shape of the circumbinary disc for binary parameters
corresponding to Kepler-16.   

The anomalous viscosity in the disc, which probably arises from MHD turbulence, is modelled using 
the standard alpha prescription for the effective kinematic viscosity $\nu=\alpha c_s H$ (Shakura \& Sunyaev 1973) 
where $H$ is the disc scale height. In this work, we use values of 
$\alpha=10^{-4}$, $10^{-3}$ and $10^{-2}$. 

The  binary semi-major axis $a_b$, eccentricity $e_b$ and mass ratio $q_b$, which remain fixed in the 
course of the simulations, are set to match the binary orbital elements obtained from Kepler 
observations. Data for Kepler-16 have been taken from Doyle et al. (2011) and
parameters for Kepler-34 and Kepler-35 have been taken from Welsh et al. (2012).
We remind the reader that for the three systems, the binary and planet parameters inferred from Kepler data 
can be found in table $1$. 

Planet evolution is initiated on a circular orbit with semi-major axis $a_p$ chosen to be
outside the disc inner cavity and in a region where the local disc eccentricity is $e_d< 0.01$.
For each disc model, we performed one simulation in which the planet mass ratio is equal to the
observed value and remains constant, and a second run in which a 20 M$_\oplus$ embryo is allowed to accrete gas from the disc. 
In the latter the onset of gas accretion occurs once migration of the low-mass embryo allows it to
reach an orbit where migration halts because the net time averaged disc torque equals zero.
Following Kley (1999), accretion by the protoplanet is modelled by removing a fraction of the 
gas located inside the Hill sphere during each time step, and this mass is added to the planet (e.g. Nelson et al. 2000; 
Pierens \& Nelson 2008). 
The gas removal rate is chosen such that the accretion timescale onto
the planet is $t_{acc}=f_{acc}t_{dyn}$, where $t_{dyn}$ is the orbital period of the planet and $f_{acc}$ is a 
free parameter.

 When calculating the torque exerted by the disc on the planet, we exclude the material contained within 
a distance $0.6R_H$ from the planet using a Heaviside filter, where $R_H=a_p(m_p/3M_\star)^{1/3}$ is the Hill radius. Such a 
value is consistent with the one that is recommended by Crida et al. (2009) to obtain an accurate estimation of the 
migration rate.
Moreover, we set the softening parameter for the gravitational potential to $b=0.4H$.
\begin{table}
\caption{Binary and planet parameters}              
\label{table1}      
\centering                                      
\begin{tabular}{c c c c}          
\hline\hline                        
 & Kepler 16 & Kepler 34 & Kepler 35 \\    
\hline                                   
 $M_A$ ($M_\odot$) & $0.69$ & $1.05$ & $0.89$\\
 $M_B$ ($M_\odot$) & $0.20$ & $1.02$ & $0.81$\\
 $m_p$ ($M_J$)     & $0.33$ & $0.22$ & $0.13$\\
 $q_b=M_A/M_B$     & $0.29$ & $0.97$ & $0.91$\\
 $q=m_p/M_*$ & $3.7\times 10^{-4}$ & $1.1\times 10^{-4}$      & $7.5\times 10^{-5}$ \\
 $a_b$ (AU) & $0.22$ & $0.23$ & $0.18$ \\      
 $a_p$ (AU) & $0.7$ & $1.09$ & $0.6$ \\
 $e_b$ & $0.16$ & $0.52$      & $0.14$ \\
 $e_p$ & $0.069$ & $0.18$ & $0.04$\\
\hline                                             
\end{tabular}
\end{table}

\subsection{Boundary conditions}

At the inner edge of the computational domain we use a closed, reflecting boundary condition. 
Because the inner boundary is located deep inside the tidally truncated cavity, there is no evidence 
of wave reflection at the boundary. In order to test the dependence of our results against the choice 
of boundary condition, we have also employed in some runs a viscous outflow boundary condition (Pierens \& Nelson 2008),
where the radial velocity in the ghost zones is set to $v_r =\beta v_r(R_{in})$, where $v_r(R_{in})=-3\nu/2R_{in}$ 
is the gas drift velocity due to viscous evolution (assuming steady mass flow) and $\beta$ is a free parameter which was set 
$\beta=5$.  At the outer radial boundary, we employ a wave-killing zone for $R>8.4$ (4.2 AU) to avoid wave reflection/excitation 
at the disc outer edge (de Val-Borro et al. 2006). The impact of this wave-killing zone on the disc shape and eccentricity is 
expected to be small since, as we will see shortly, the disc is almost circular well inside the wave-killing zone (see Fig. \ref{eccr} below). 

\section{Evolution of the binary-disc system}
\subsection{Evolution of the disc eccentricity}
\begin{figure*}
\includegraphics[width=0.49\textwidth]{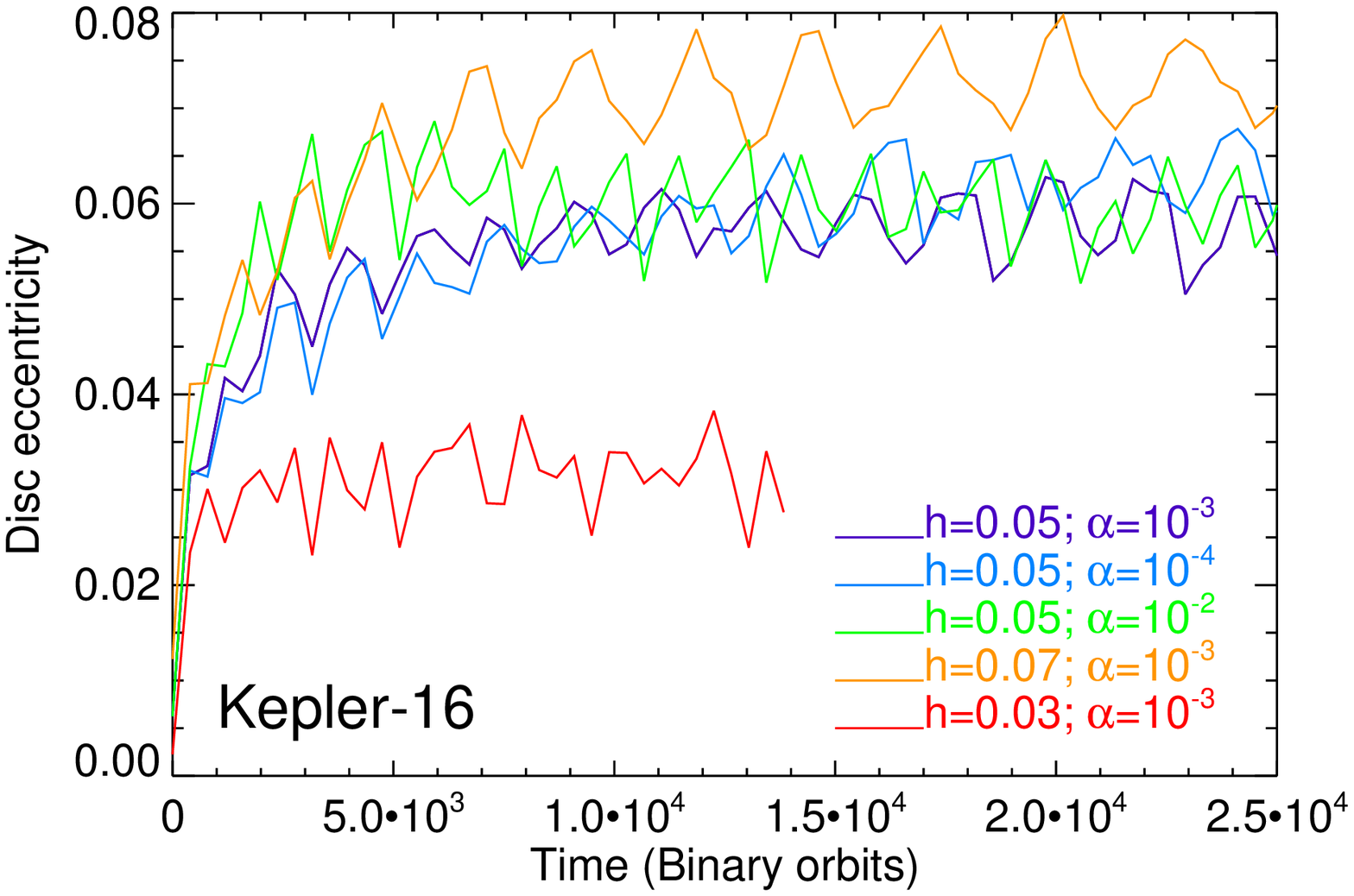}
\includegraphics[width=0.49\textwidth]{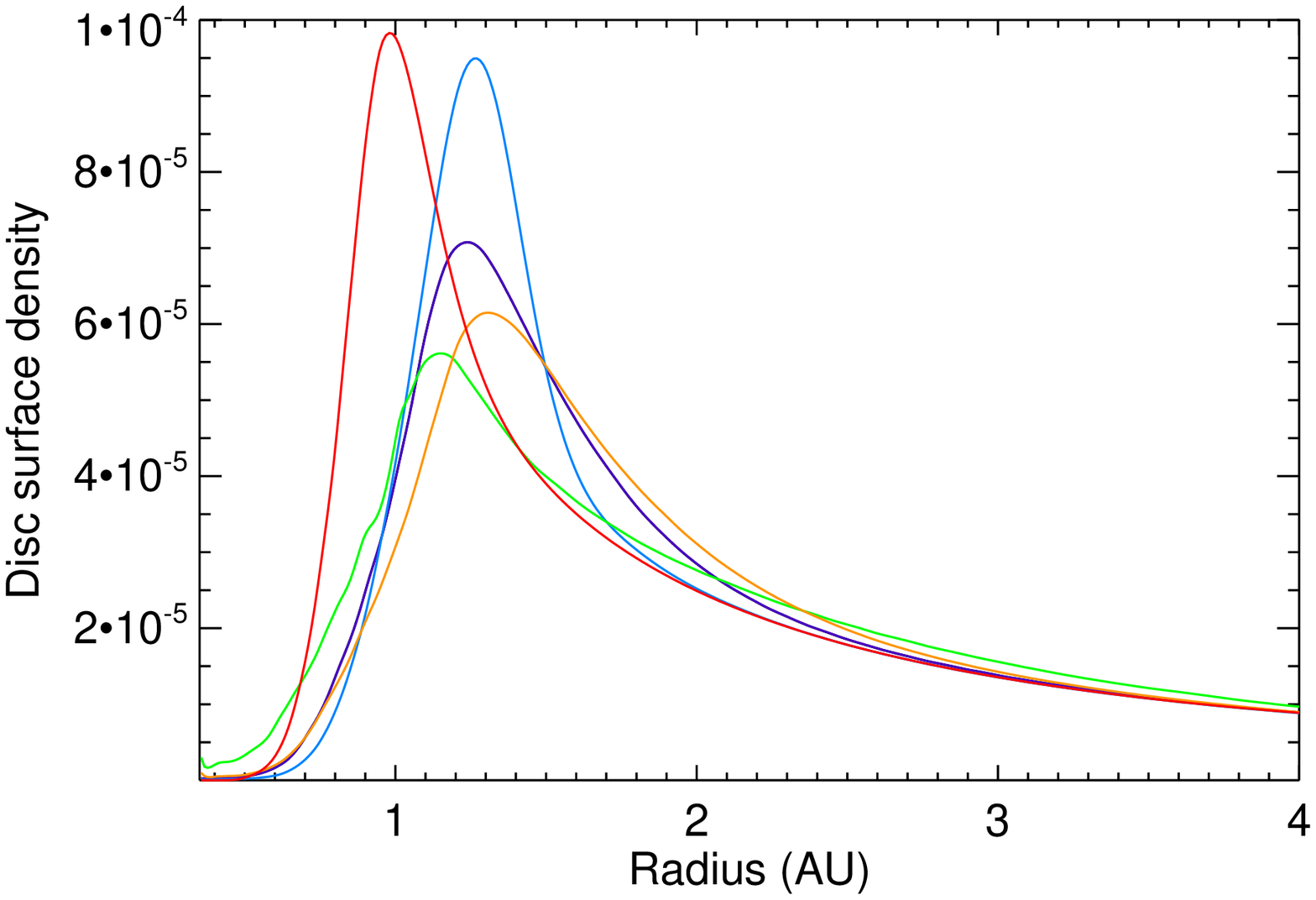}
\includegraphics[width=0.49\textwidth]{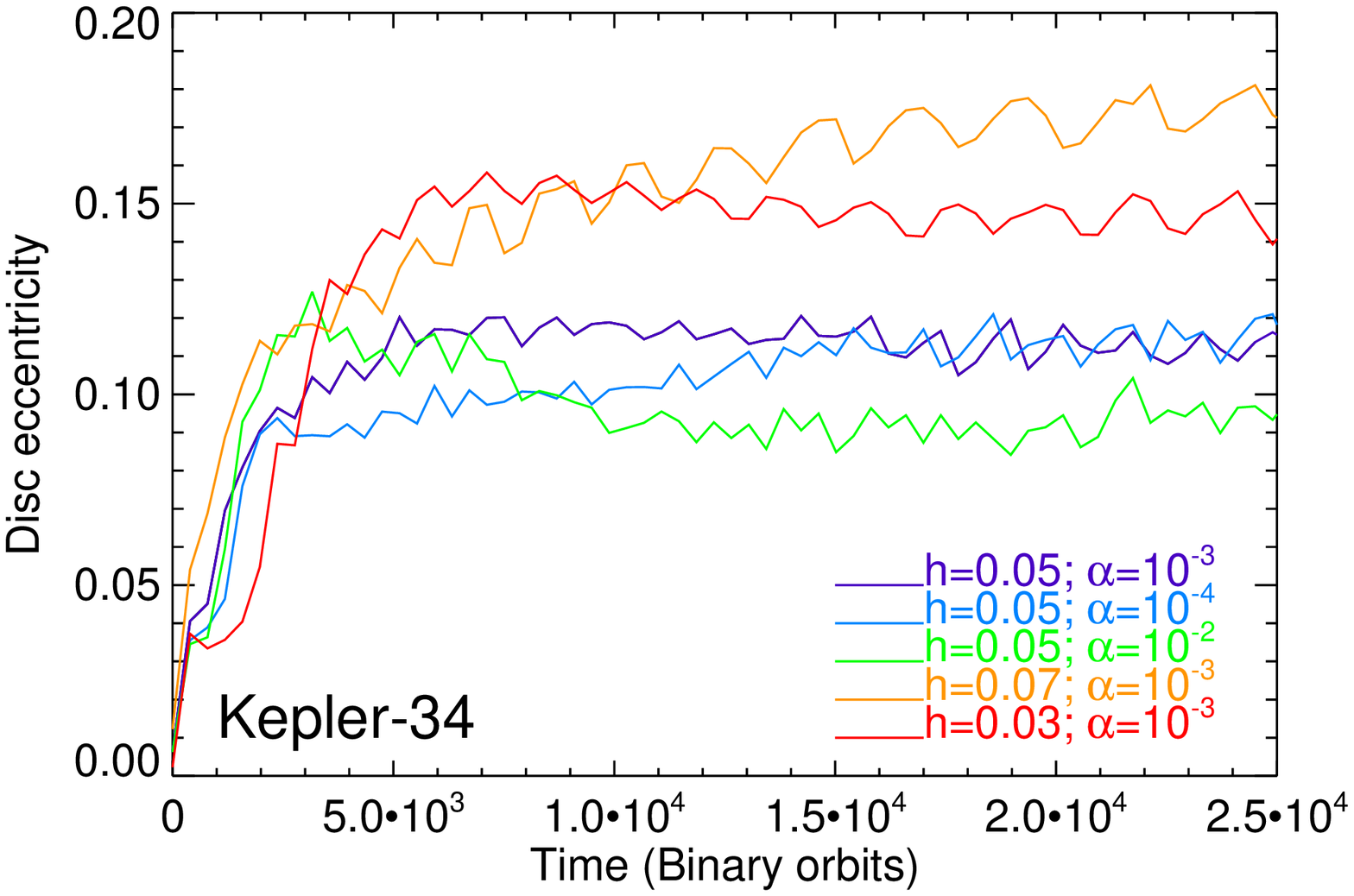}
\includegraphics[width=0.49\textwidth]{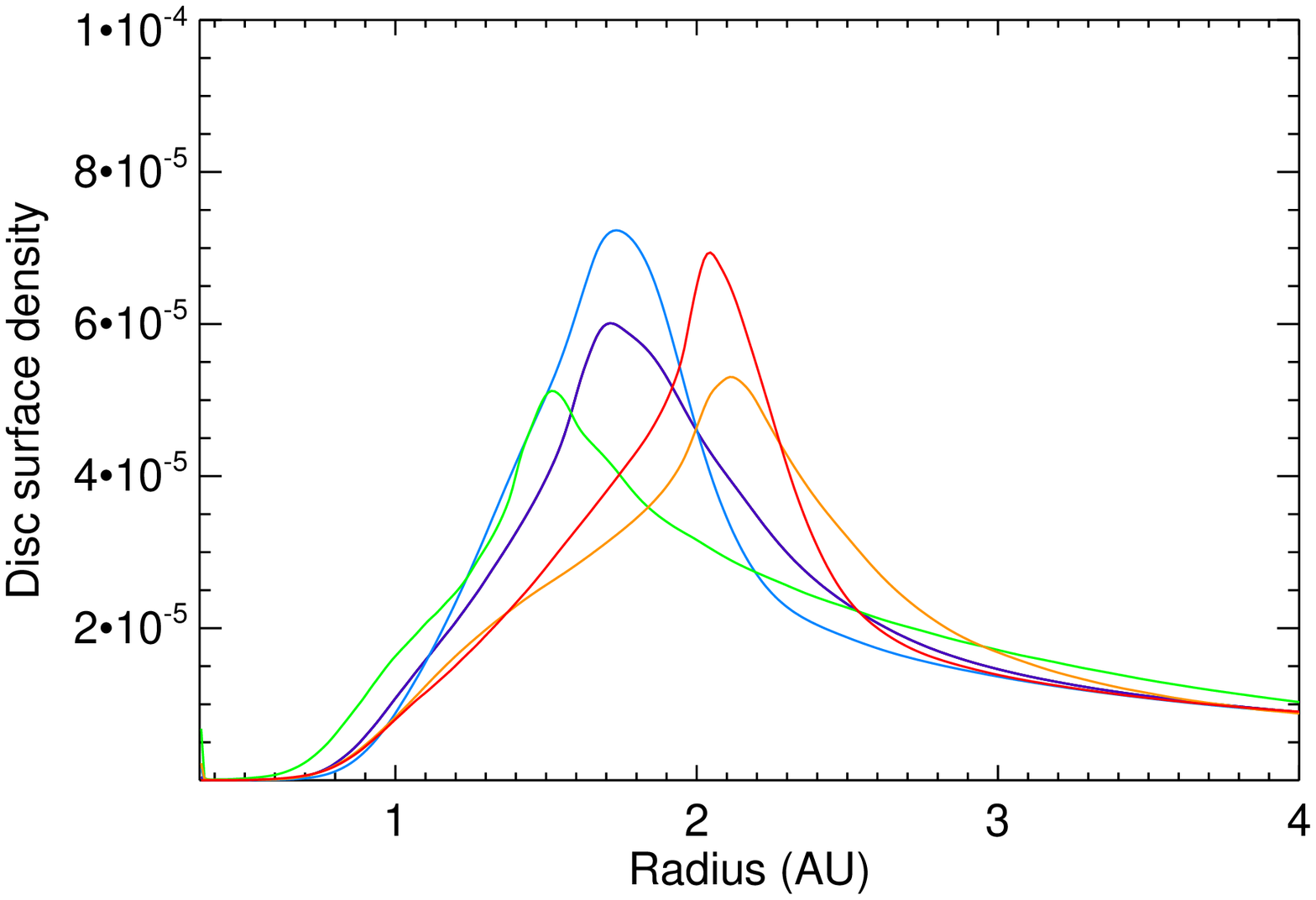}
\includegraphics[width=0.49\textwidth]{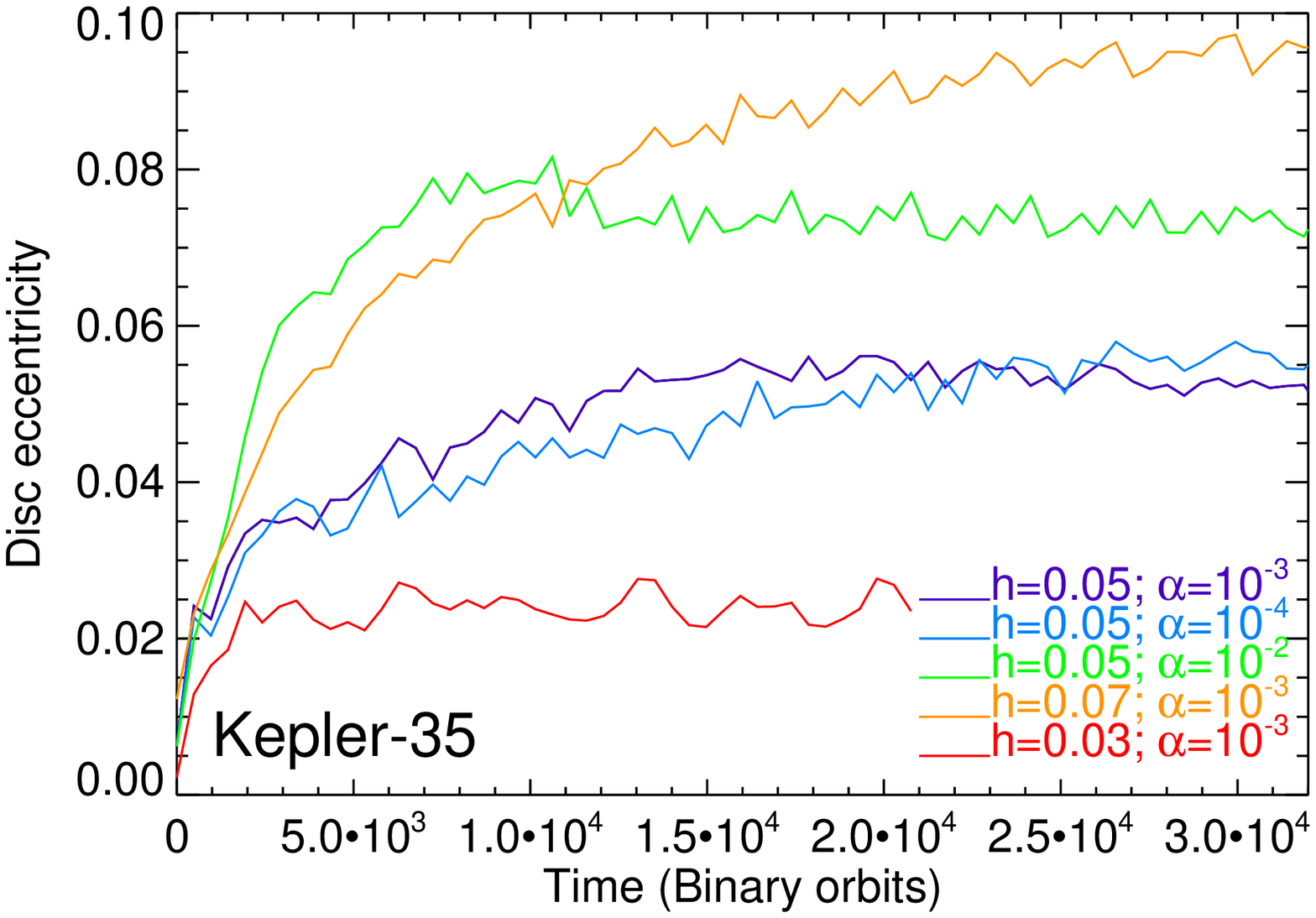}
\includegraphics[width=0.49\textwidth]{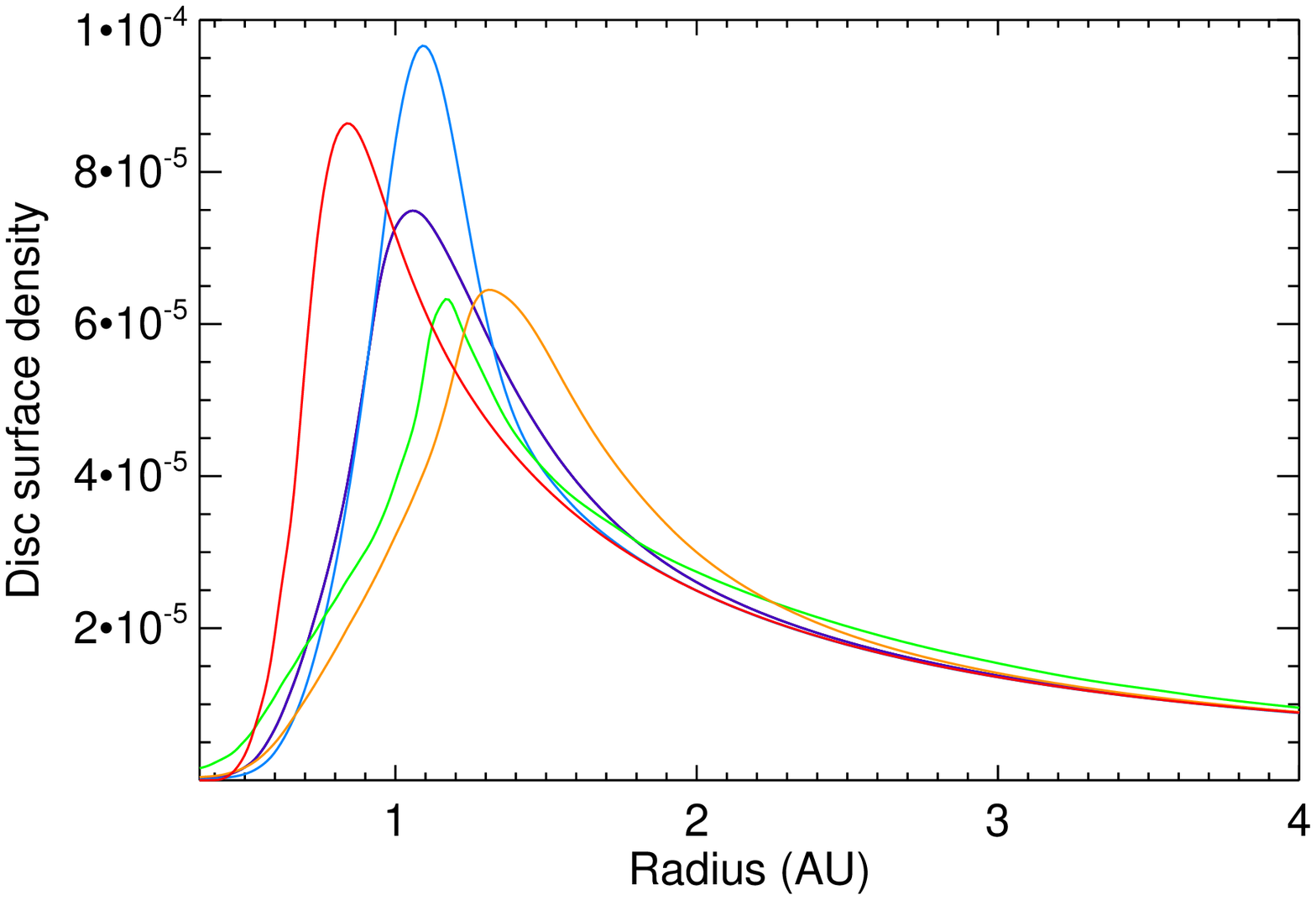}
\caption{{\it Left panel:} Time evolution of the disc eccentricity for Kepler-16, Kepler-34 and 
Kepler-35. {\it Right panel:} Disc surface density profile at quasi-steady state for the three systems.  }
\label{eccentricity}
\end{figure*}
Before presenting simulations of the evolution of embedded circumbinary planets, we first 
examine briefly the evolution of the binary-disc system and the growth of disc eccentricity. 
This is not intended to be an exhaustive analysis of this issue as our primary concern
in this paper is the evolution of circumbinary planets. We will present a more detailed
analysis of the evolution of circumbinary discs in a future publication, where we will consider
the influence of radiative processes, self-gravity, binary mass ratios and binary eccentricities.

\begin{figure}
\includegraphics[width=\columnwidth]{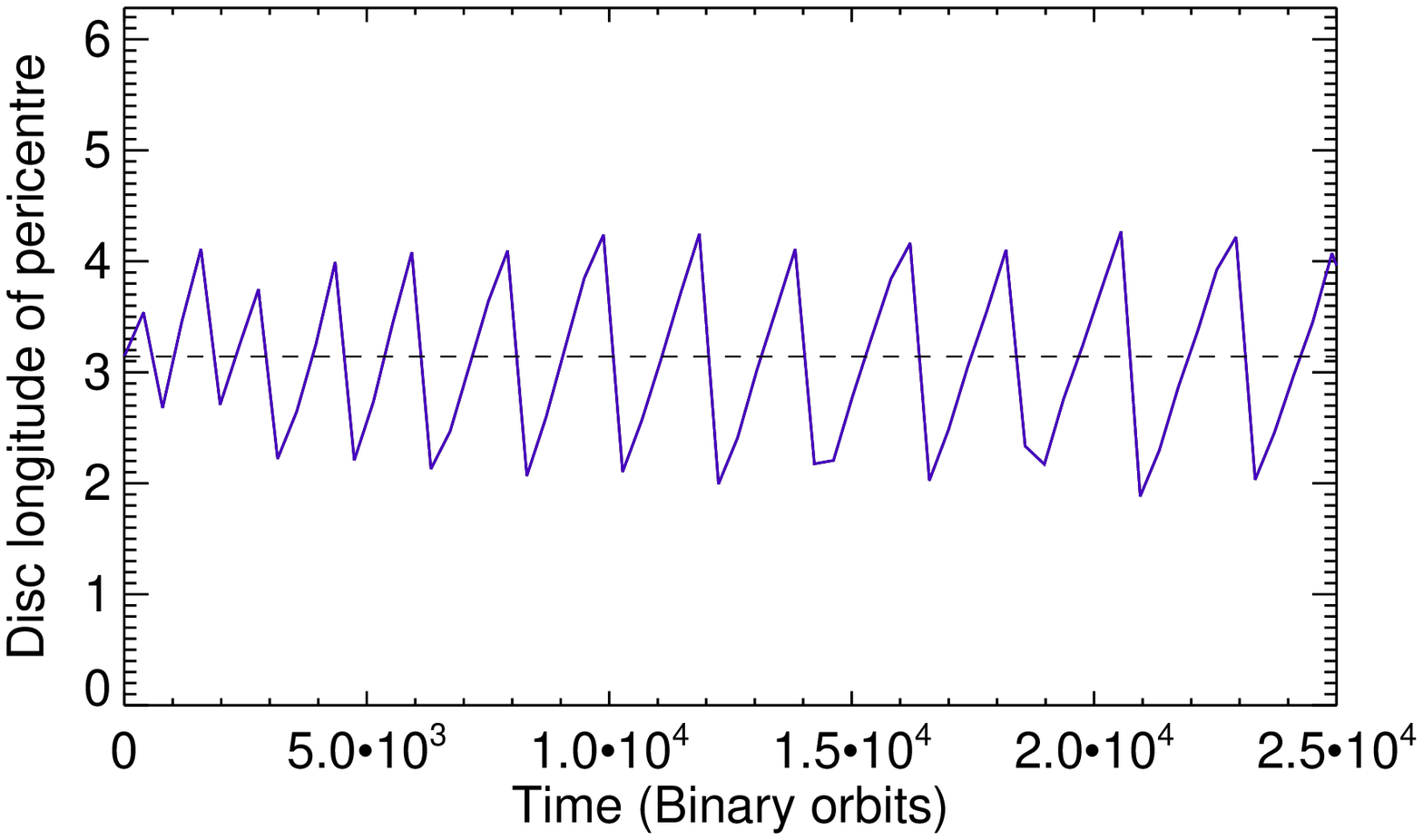}
\caption{Time evolution of the disc longitude of pericentre for the disc model with $h=0.05$ and 
$\alpha=10^{-3}$. The dashed line corresponds to the longitude of pericentre of the binary.} 
\label{wdisc}
\end{figure}

The left panel of Fig. \ref{eccentricity} shows the time evolution of the disc eccentricity $e_d$ for the 
three systems, where $e_d$ is defined as (Pierens \& Nelson 2007):
\begin{equation}
e_d=\frac{\int_0^{2\pi}\int_{R_{in}}^{R_{out}}\Sigma e_c dS}{\int_0^{2\pi}\int_{R_{in}}^{R_{out}}\Sigma dS}.
\end{equation}
In the previous equation,  $e_c$ is the eccentricity of a fluid element computed at the center of each grid cell 
and $dS$ the surface area of one grid cell.
For each run, the disc eccentricity initially increases and then reaches a saturated value once viscous damping 
equilibrates the eccentricity forcing rate due to the binary. The observed oscillations in the disc 
eccentricity arise because the binary longitude of pericentre maintains a constant value while the disc librates 
with a libration amplitude of $\sim 108$ degrees. This is illustrated in Fig. \ref{wdisc} which shows the time 
evolution of the disc longitude of pericentre $\omega_d$ for the Kepler-16 run with $\alpha=10^{-3}$ and $h=0.05$, 
where $\omega_d$ is defined by:
\begin{equation}
\omega_d=\frac{\int_0^{2\pi}\int_{R_{in}}^{R_{out}}\Sigma \omega_c dS}{\int_0^{2\pi}\int_{R_{in}}^{R_{out}}\Sigma dS}.
\end{equation}
$\omega_c$ is the longitude of pericentre of a fluid element evaluated at the center of each grid cell. The 
fact that the eccentricity oscillation amplitudes are higher in models with $h=0.07$ is likely due to pressure effects which tend 
to reduce the prograde precession rate of the disc. 

We observe a clear tendency for the final value of $e_d$ to increase with increasing $h$.
The Kepler-34 run with $h=0.03$ is the one anomaly, which can perhaps be explained by the smaller
viscous stress leading to weaker damping of the locally generated eccentricity at large radius.
We leave further investigation of this issue for the future publication mentioned at the beginning of this section.

Comparing the upper and lower panels of Fig. \ref{eccentricity} for Kepler-16 and Kepler-35, where the binaries 
have similar eccentricities but different mass ratios, reveals that the saturated value of the disc eccentricity is 
almost independent of the binary mass ratio. For example, for our fiducial disc model with $h=0.05$ and $\alpha=10^{-3}$, 
the steady state value for the disc eccentricity is $e_d\sim 0.55$ in the Kepler-16 run while $e_d\sim 0.5$ for Kepler-35. 
Comparing the middle and lower panels for Kepler-34 and Kepler-35, however,  where the binaries have
similar values for $q_b$ but very different eccentricities, demonstrates that the final value for $e_d$ strongly depends 
on the binary eccentricity. Not surprisingly, it is found to be higher in Kepler-34 runs with, for example, a stationary value of 
$e_d\sim 0.11$ reached in the fiducial run with $h=0.05$ and $\alpha=10^{-3}$. Given that the value for the disc eccentricity is 
closely related to the size of the inner cavity, as we will see shortly, this is in good agreement with the results of 
Artymowicz \& Lubow (1994) which suggest that the gap size is strongly dependent on the binary eccentricity but depends weakly 
on the binary mass ratio.\\

An issue that needs to be investigated is how the final value for the disc eccentricity compares with the 
expected forced eccentricity resulting from the secular interaction with the binary. The forced eccentricity $e_f$
at radius $R$ is given by (Moriwaki \& Nakagawa 2004):
\begin{equation}
e_f=\frac{5}{4}\frac{M_A-M_B}{M_\star}\frac{a_b}{R}\frac{1+3e_b^2/4}{1+3e_b^2/2}.
\end{equation}
For the Kepler-16 system, this gives $e_f\sim 0.02$ at $1$ AU while for Kepler-34 and 
Kepler-35 which have mass ratios close to unity, the forced eccentricity is much lower 
($e_f\sim 10^{-3}$ at $1$ AU). Except for the Kepler-16 run with $h=0.03$ and $\alpha=10^{-3}$, it 
therefore appears that the saturated value for $e_d$ is generally much higher than the forced eccentricity, 
which is in good agreement with the results of Pelupessy \& Portegies Zwart (2013). 
This indicates that the contribution to the disc eccentricity growth from secular interaction is minimal. 

The right panel of Fig. \ref{eccentricity} displays the steady-state disc surface density profile for all calculations,
time-averaged over $3\times 10 ^3$ binary orbits. As the values for $h$ and $\alpha$ are 
decreased the inner cavity deepens and the cavity edge steepens.  Not surprisingly there also appears a general trend 
for the location of the density peak $R_{gap}$ to scale with the disc eccentricity. Here, we find that 
$R_{gap}$ can be reasonably well approximated by the following fitting relation:
\begin{equation}
R_{gap}\sim (3.8+35e_d)a_b.
\end{equation}

\subsection{Origin of the disc eccentricity}

\begin{figure}
\includegraphics[width=\columnwidth]{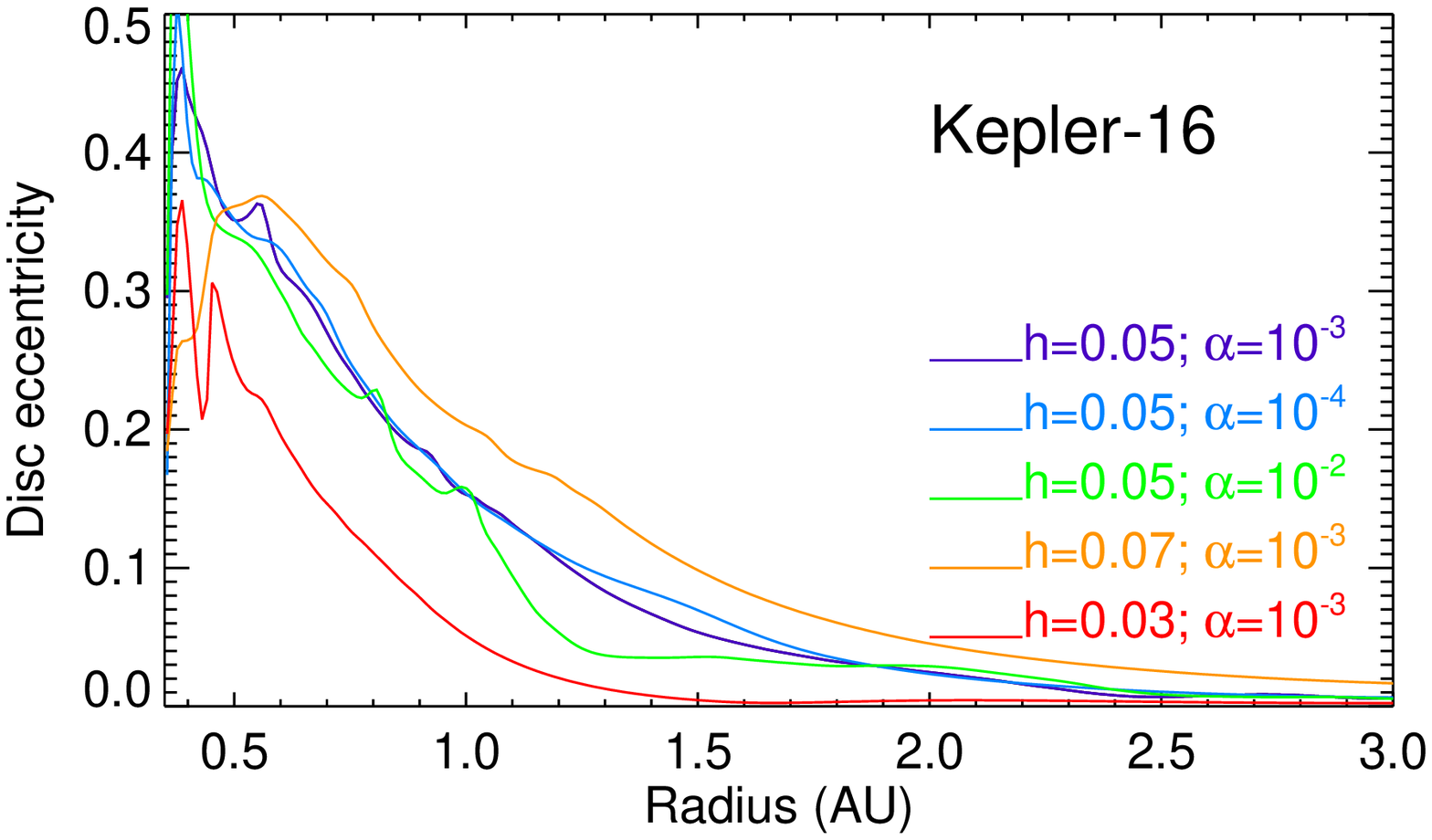}
\includegraphics[width=\columnwidth]{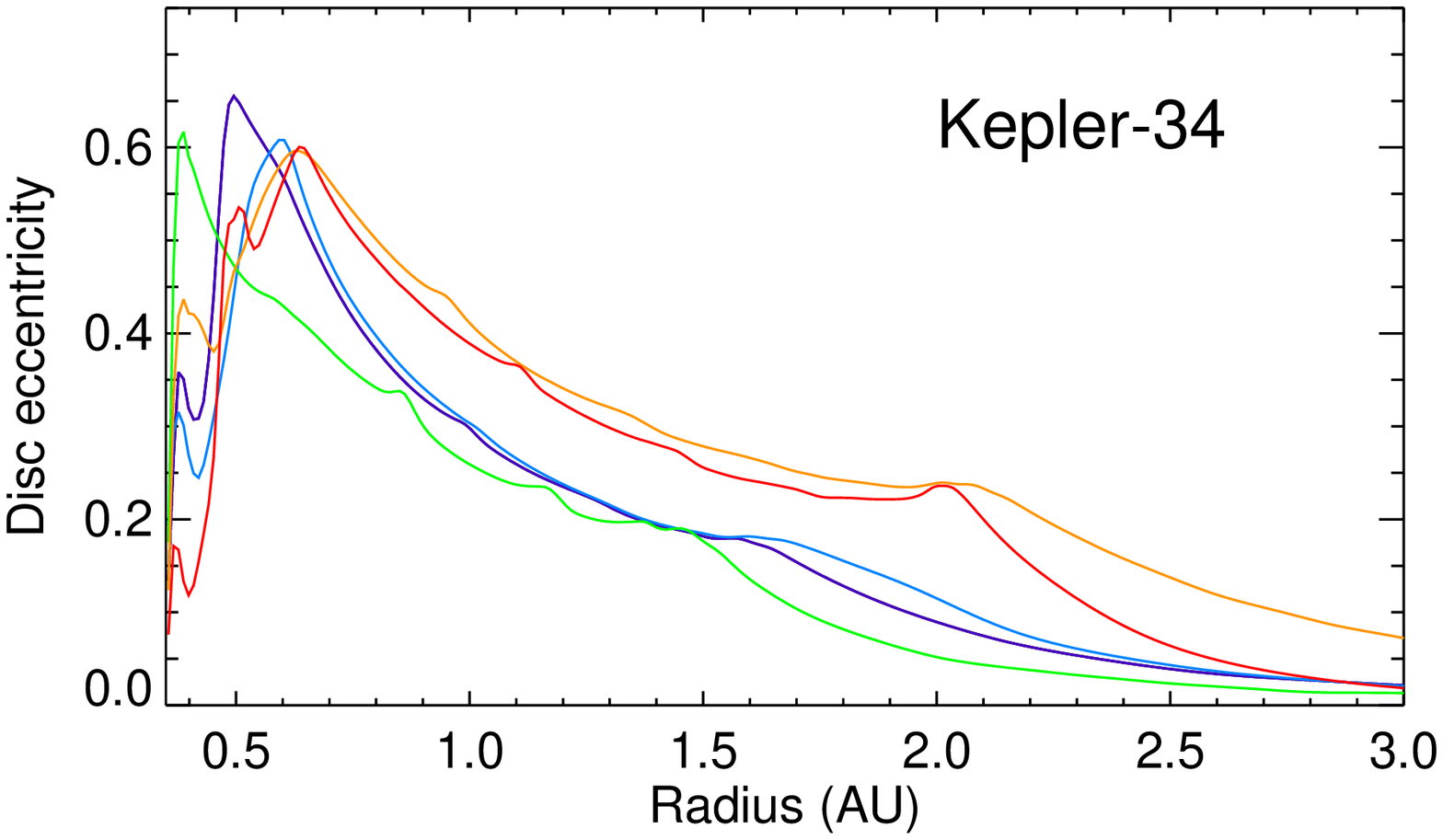}
\includegraphics[width=\columnwidth]{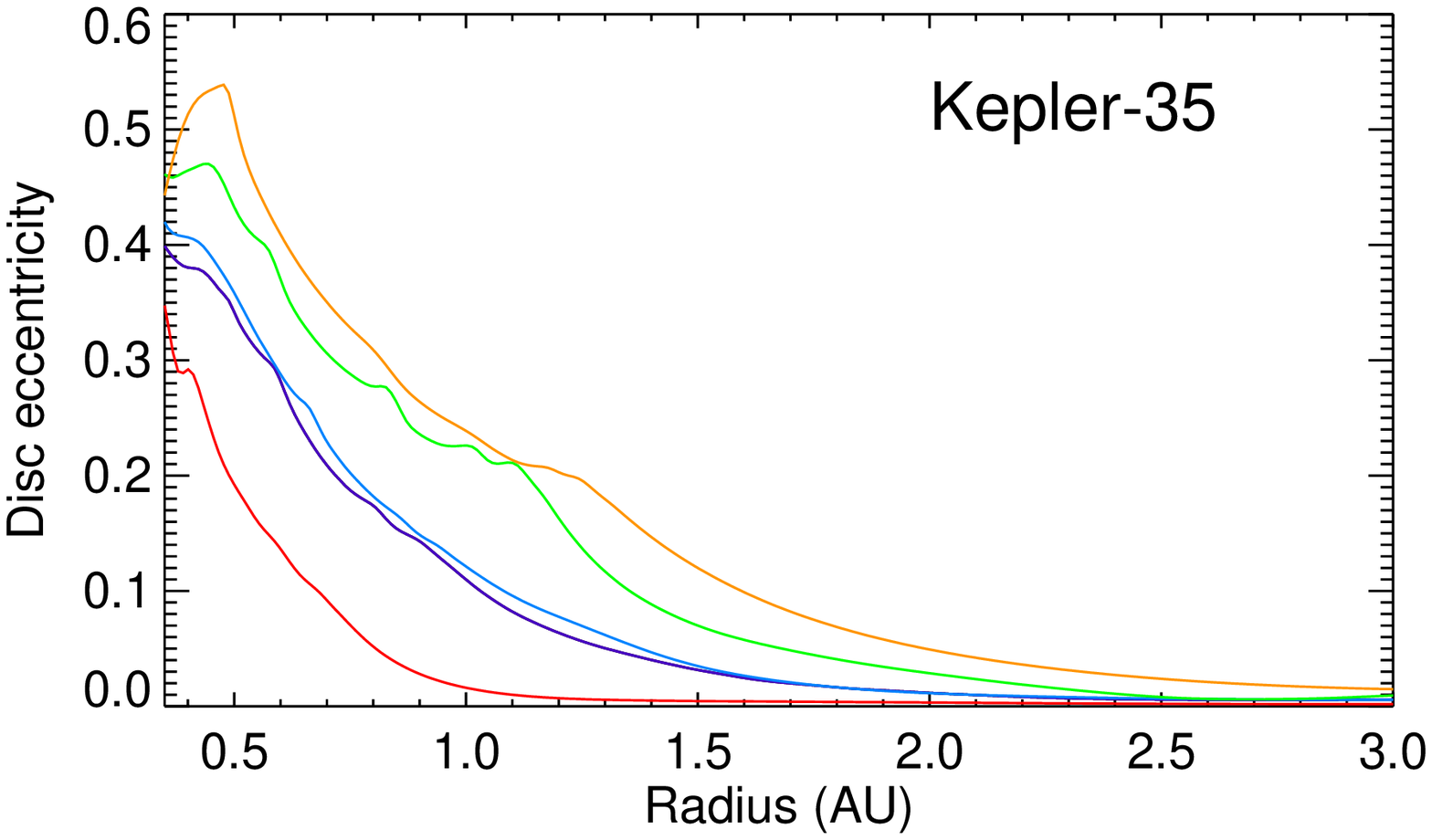}
\caption{Radial distribution of disc eccentricity for the Kepler-16, Kepler-34 and Kepler-35 runs}
\label{eccr}
\end{figure}


Previous studies of circumbinary discs (e.g. Papaloizou et al. 2001; Pierens \& Nelson 2007) 
have shown that interaction with the binary can cause the disc to become eccentric. 
As we have discussed above, the forced eccentricity of the disc due to secular interaction
with the binary should be small. For a circular binary, Papaloizou et al. (2001) have 
shown that the origin of the disc eccentricity involves a parametric instability that causes a
spiral wave to be launched at the $3:1$ eccentric Lindblad resonance. This instability is generated by non-linear 
mode coupling between an initial $m=1$ eccentric disturbance in the disc and the $m=1$ component of the binary 
potential. This produces a $m=2$ wave with pattern speed $\Omega_b/2$ that is launched at the $3:1$ Lindblad resonance.
The wave transports angular momentum outward, causing the growth of disc eccentricity (Papaloizou et al. 2001). 
The mechanism at play is very similar to that described by Lubow (1991) in application to superhump
phenomena in close binary systems where the disc orbits one component of the binary system.
In this case a small binary mass ratio ($q_b\sim 0.1$) allows the $1:3$ Lindblad resonance to be present
in the disc, leading to the growth of disc eccentricity through a similar instability. 

To explore the origin of the disc eccentricity observed in Figs. 1 and 3 we have conducted a
number of test calculations. The aim of this study is simply to explore the primary
influences on the growth of the eccentricity. In a future paper we will provide an in-depth
analysis of disc eccentricity growth in circumbinary systems.
The findings of our modest present study may be summarised as follows:
\begin{itemize}
\item Simulations of circumbinary discs with binary mass ratios equal to those in the
Kepler-16 and 35 systems typically show the growth of substantial disc eccentricity when 
the binary eccentricity is set to zero. This clearly demonstrates that nonlinear mode
coupling and the 3:1 Lindblad resonance play an important role.
\item Simulations were performed for circular binary systems where the inner boundary of
the computational domain was placed outside the location of the 3:1 resonance. Disc eccentricity
growth was not observed in these cases, confirming the conclusion in the previous paragraph.
\item Simulations were performed for binary systems similar to Kepler-34 with the inner
boundary of the computational domain placed between the locations of the 3:1 and 4:1
Lindblad resonance. Strong disc eccentricity growth was observed in this case, suggesting
that nonlinear mode coupling involving higher-order terms in the potential of the
binary also cause the growth of disc eccentricity, at least for the highly eccentric
Kepler-34 binary system.
\end{itemize}

When both the disc eccentricity and surface density profile have reached a quasi steady-state, we include a planet 
that interacts gravitationally with both the binary and the circumbinary disc. This planet initially evolves on a 
circular orbit at $a_p=1.5$ AU in Kepler-16 and Kepler-35 runs while the initial semi-major axis
$a_p=2$ AU in Kepler-34 runs. For the three systems, we tested two different scenarios:\\
(i) We first considered the case of a fully-formed planet for which the planet mass is set to the observed value.
For Kepler-16b, this corresponds to a mass ratio $q=3.7\times 10^{-4}$, and for Kepler-34b and Kepler-35b $q=1.1\times 10^{-4}$ and 
$q=7.5\times 10^{-5}$, respectively (see Table $1$). \\
(ii) We then performed a series of simulations in which the planet mass ratio is $q=3\times10^{-5}$. In that case, 
the interaction with the disc is linear and we expect the planet to undergo Type I migration. A typical 
observed evolution outcome of such simulations is stopping of inward migration close to the tidally 
truncated cavity. Once the planet has reached a quasi-stationary orbit, we allow gas accretion onto the core 
and examine whether such a scenario can reasonably reproduce the observed systems. Removal of the gas disc
occurs in these simulations to mimick photoevaporation on a specified time scale.

\section{Evolution of fully-formed planets}
\label{sec:formed}
\begin{figure*}
\includegraphics[width=\textwidth]{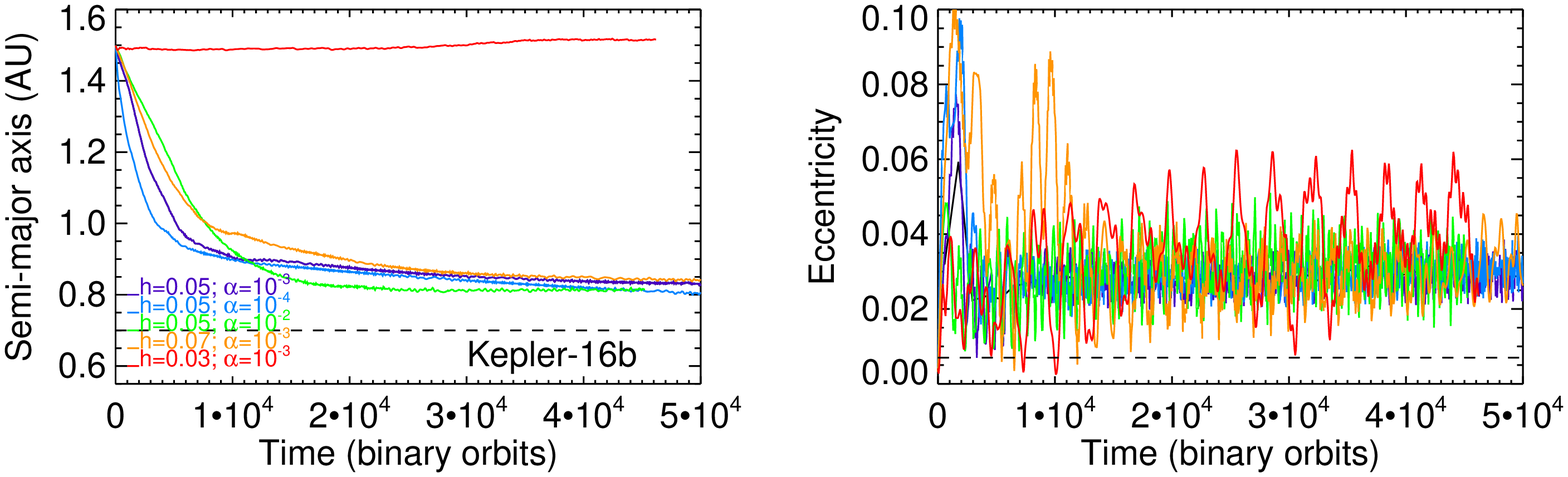}
\includegraphics[width=\textwidth]{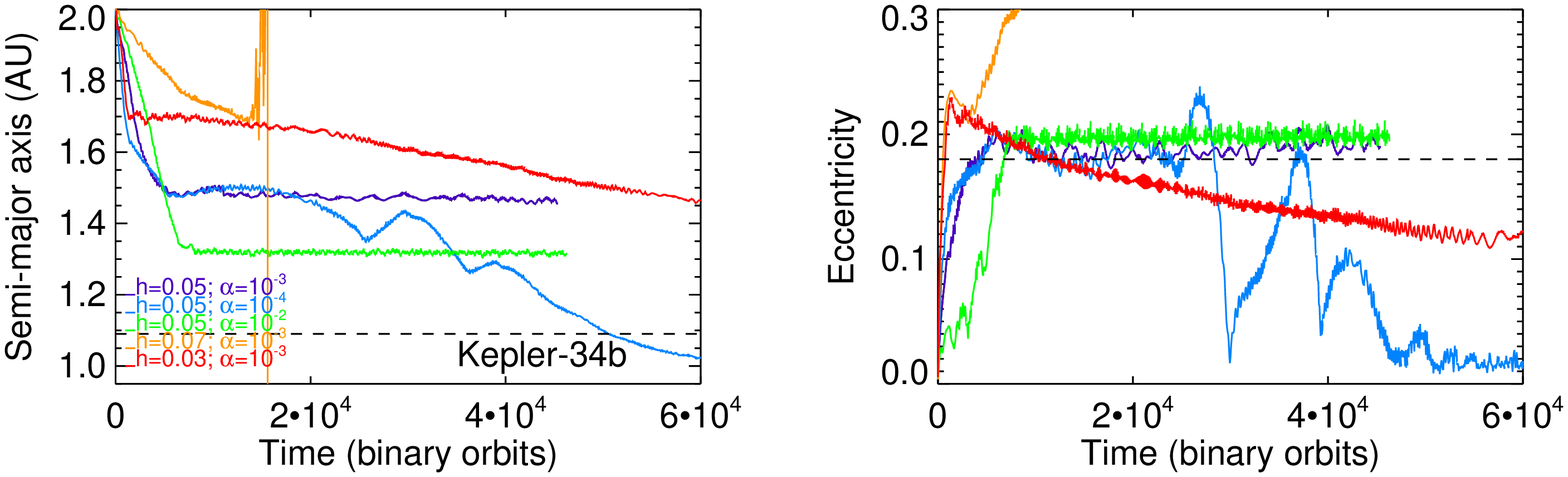}
\includegraphics[width=\textwidth]{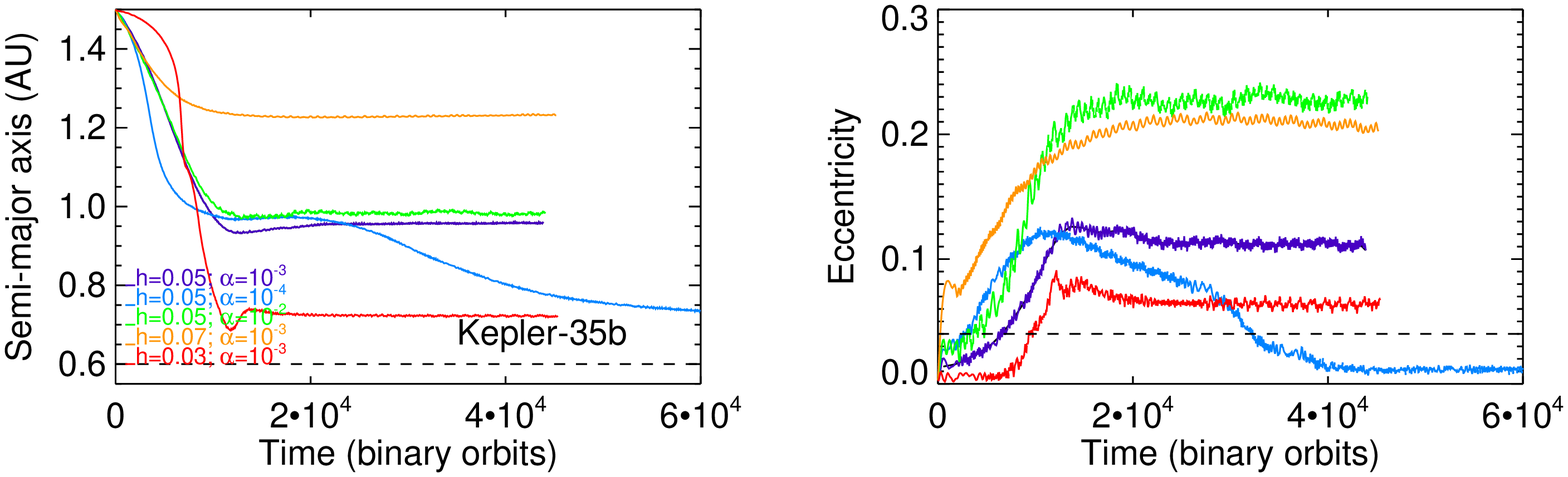}
\caption{{\it Left panel:} Time evolution of the semi-major axis of an embedded planet with, 
from top to bottom, mass ratio corresponding to that of Kepler-16b, Kepler-34b and Kepler-35b . {\it Right panel:} Time 
evolution of the eccentricity. For each parameter, the dashed line correspond
to the observed value}
\label{mf}
\end{figure*}
Here we examine the orbital evolution of embedded planets with masses corresponding to the observed values for the three 
systems Kepler-16, Kepler-34 and Kepler-35. The aim is to investigate how the evolution depends on the employed disc model,
and whether or not decent agreement with the observed planet parameters can be obtained.
\subsection{Kepler-16b}
\begin{figure}
\includegraphics[width=\columnwidth]{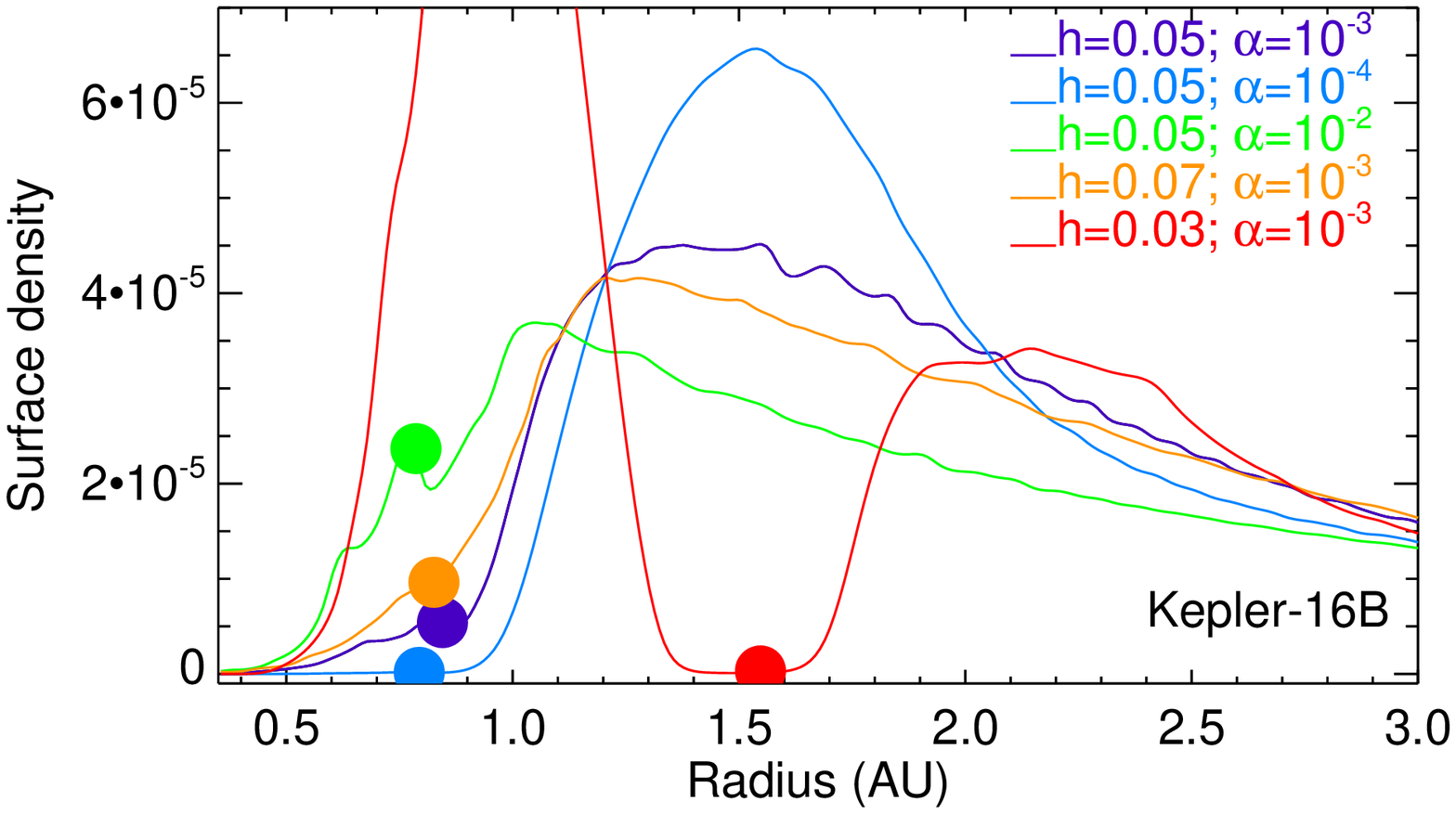}
\includegraphics[width=\columnwidth]{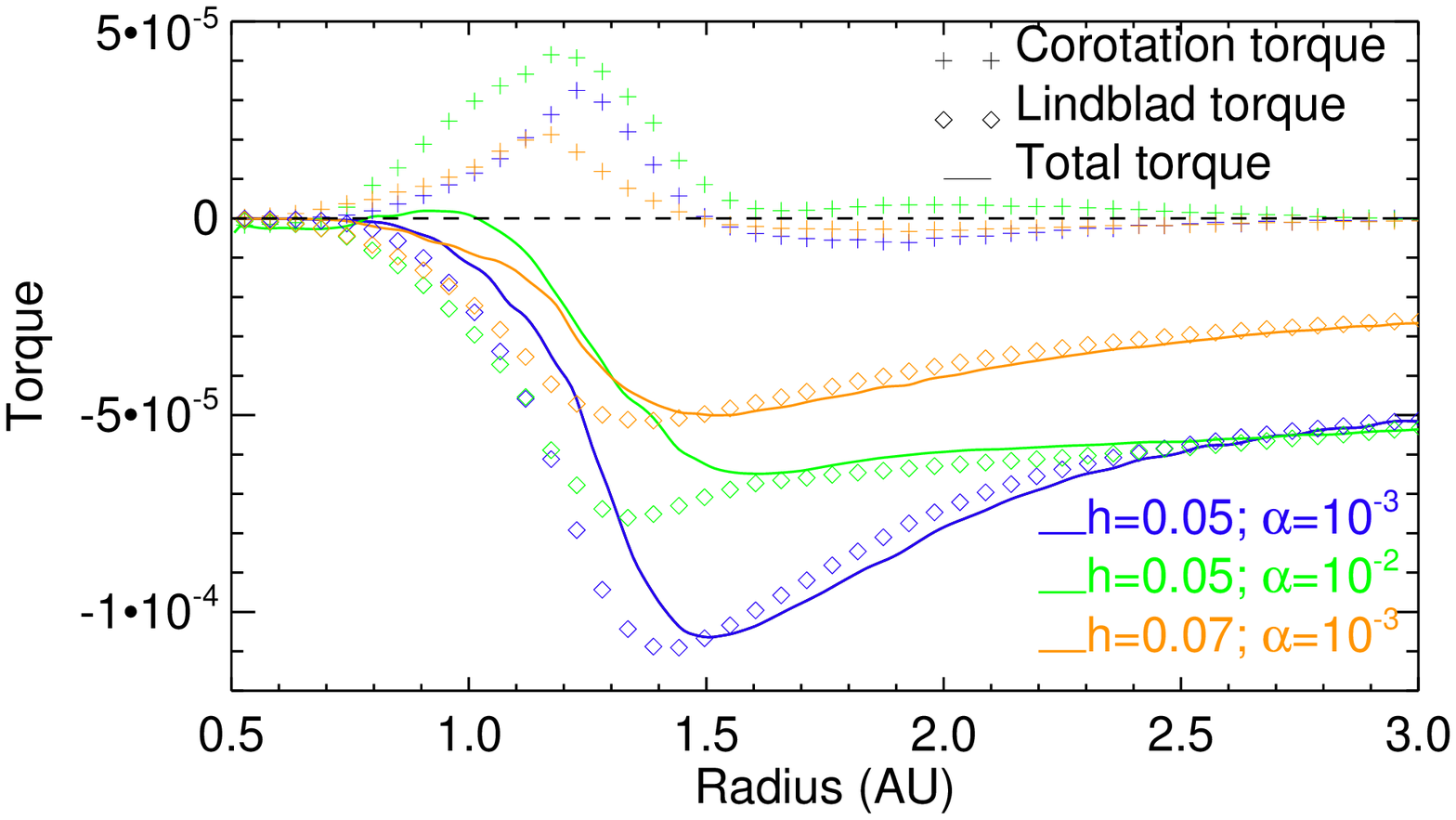}
\caption{ {\it Upper panel:} Surface density profile at $t=4\times 10^4$ binary orbits for the 
Kepler-16b runs. {\it Lower panel:} Semi-analytical torques exerted on Kepler-16b as a function of the orbital 
radius.}
\label{torque-kep16}
\end{figure}

For our different isothermal disc models, the orbital evolution of an embedded planet with mass ratio corresponding to that 
of Kepler-16b ($q=3.7\times 10^{-4}$) is presented in the upper panel of Fig. \ref{mf}. 
Using the gap opening criteria of Crida et al. (2006) which predicts that gap opening should occur provided that 
\begin{equation}
1.1\left(\frac{q}{h^3}\right)^{-1/3}+\frac{50\alpha h^2}{q}\le 1
\label{gapopening}
\end{equation}
we expect an embedded circumbinary planet with mass ratio $q=3.7\times 10^{-4}$ to significantly perturb the 
underlying surface density profile in the discs with $h=0.03$, $h=0.05$ and $\alpha\le 10^{-3}$.
This is illustrated in the upper panel of Fig. \ref{torque-kep16} which displays the disc surface 
density profile at $t=4\times 10^4$ orbits for all disc models. Gap formation is clearly apparent
in the case with $h=0.03$, leading also to a substantial increase of the surface density in the inner disc. This results 
in a strong positive torque being exerted by the inner disc on the planet that prevents inward migration.
This result is robust against the choice of inner boundary condition as we performed a simulation using a viscous outflow 
condition (see Pierens \& Nelson 2008) and found no differences. 

 For disc models with $h \ge 0.05$,  early evolution typically involves rapid inward migration due to the high surface density at the outer edge of the tidally truncated cavity. At later times the migration 
rate decreases as the planet enters the cavity and evolves in a region where the surface density gradient is positive. 
During the run with $\alpha=10^{-2}$ and $h=0.05$ inward migration stops with the planet settling into an orbit with 
eccentricity $e_p\sim 0.04$. Migration appears to stop due to the development of a strong positive corotation torque which
counterbalances the Lindblad torque within the inner cavity where the surface density has a positive gradient (Masset et al. 2006a), 
as illustrated in the upper panel of Fig. \ref{torque-kep16}. Saturation of the corotation torque can be prevented provided 
the viscous timescale across the horseshoe region $t_{visc}\sim x_s^2/3\nu$, where $x_s$ is the 
half-width of the horseshoe region, is smaller than the horseshoe libration timescale $\tau_{lib}\sim 8\pi a_p/(3\Omega_p x_s)$. 
Equating these time scales yields an estimate of the minimum value for the viscous stress parameter, $\alpha$, required to avoid 
corotation torque saturation. We find:
\begin{equation}
\alpha\sim 1.3\times 10^{-2} q h^{-2}
\end{equation}
where we have used the fact that for $q> 10^{-4}$, the half-width of the horseshoe region is $x_s\sim R_H$ (Masset et al. 2006b). 
For $h=0.05$ this gives $\alpha\sim 2\times 10^{-3}$, indicating that the corotation torque is indeed unsaturated for this model. 
In order to clearly demonstrate that this effect is at work here, we have followed the approach of Pierens \& Nelson (2007) and 
computed semi-analytically the torques experienced by the planet using the formulae in Paardekooper et al. (2011). The 
corotation, Lindblad and total torques exerted on the planet for the three models with $h\ge0.05$ and 
$\alpha\ge 10^{-3}$ are presented as a function of the orbital radius in the  
lower panel of Fig. \ref{torque-kep16}. Only in the model with $h=0.05$ and $\alpha=10^{-2}$ is the corotation 
torque seen to  be larger than the Lindblad torque inside the cavity, giving rise to the location
$R\sim 1$ AU where the total torque cancels. Such a value is consistent with the simulation result that 
migration stops at $R\sim 0.82$ AU. 

For the two models with $h=0.05$,  $h=0.07$ and $\alpha=10^{-3}$ the corotation torque is much more saturated and the 
planet is expected to migrate inward further. Consistent with this prediction, the planet is observed to migrate 
all the way into the inner cavity in the simulation with $h=0.07$. Due to the very long run times required by the simulations, 
the evolution outcome remains uncertain but the fact that there is a significant amount of gas within the cavity for this disc model 
(see upper right panel of Fig. \ref{eccentricity}) suggests that inward migration may proceed until the planet enters the 
region of dynamical instability located at $R\le 0.64$ AU. Extrapolating the observed migration rate forward in time, we estimate 
that the planet will migrate from $R=0.83$ AU to $R=0.64$ AU in $\sim 2\times 10^4$ yr, which is much smaller than the disc life time. 
An alternative outcome, however, may be capture in a mean motion resonance with the binary before dynamical instability occurs. 

\begin{figure}
\includegraphics[width=\columnwidth]{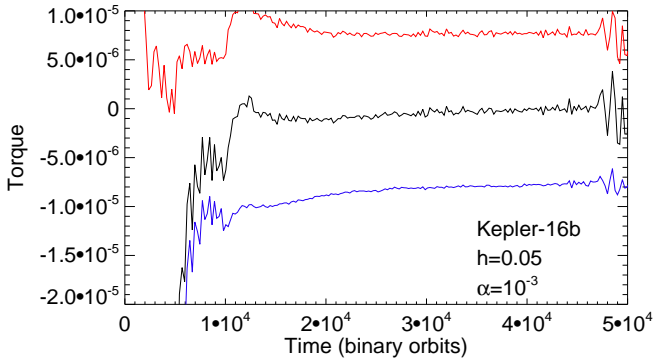}
\caption{Time evolution of the inner (red), outer (blue) and total (black) torque exerted on Kepler-16b for a 
disc model with $h=0.05$ and $\alpha=10^{-4}$.}
\label{torque-h5a3-kep16}
\end{figure}

Contrary to the above prediction concerning saturation of the corotation torque, the run with $h=0.05$ and $\alpha=10^{-3}$ 
shows stalling of inward migration from $t\sim 4\times 10^4$ P. This stalling does not arise because of a strong positive 
corotation torque, but rather because migration and gap formation by the planet, combined with tidal truncation by the binary,
cause the disc material sitting between the planet and cavity to form a narrow ring. This increases the local surface density in the ring, 
giving a stronger positive contribution to the Lindblad torque experienced by the planet. 
This is illustrated in Fig. \ref{torque-h5a3-kep16} which shows the evolution of the total torque experienced by the planet, and the
contributions from the disc lying interior and exterior to the planet semi-major axis. As time proceeds the amplitude of the (positive) 
inner-disc torque increases while the (negative) outer-disc torque weakens and the planet halts when the inner and outer torque are in balance. 

\begin{figure}
\includegraphics[width=\columnwidth]{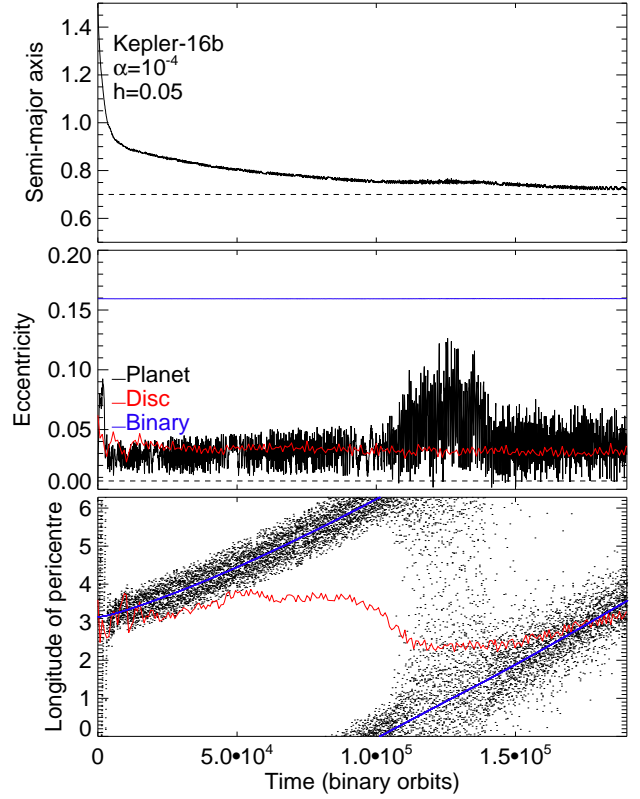}
\caption{{\it Upper panel:} Time evolution of the planet semi-major-axis for the Kepler-16b run with $\alpha=10^{-4}$ and $h=0.05$. 
{\it Middle panel:} Time evolution of the planet and disc eccentricities. {\it Lower panel:} Time evolution of the planet and disc 
longitudes of pericentre.} 
\label{h5a4-kep16}
\end{figure}

For the disc model with $h=0.05$ and $\alpha=10^{-4}$, a similar process does not occur simply because the inner cavity is much 
deeper in that case. As can be shown in the upper panel of Fig. \ref{torque-kep16}, the inner disc is completely depleted 
and the evolution outcome appears to be inward migration until the amount of local gas in the cavity becomes 
too small to make the planet migrate further in. For this model we have performed a long term evolution run for $\sim 1.5\times 10^5$ P 
and the results are presented in  Fig. \ref{h5a4-kep16}. At $t\sim 10^5$ P, we see that inward migration is 
temporarily halted as the planet reaches $a_p\sim 0.74$ AU. This corresponds to the location of the $6:1$ 
resonance with the binary. From this time the planet evolves for $\sim 5\times 10^4$ binary orbits without migrating 
just outside the $6:1$ resonance with its eccentricity oscillating between $0.007\le e_p\le 0.1$.  At $t\sim 1.4\times 10^5$ P 
the planet migrates inward through the 6:1 resonance and halts at semi-major axis close to the currently observed value ($a_p\sim 0.7$ AU). 
At the end of the simulation, the planet eccentricity oscillates between $0.005 \le e_p \le 0.07 $, while Kepler data gives $e_p\sim 0.0069$.   
 

\subsection{Kepler-34b}
\label{sec:kepler34}

\begin{figure}
\includegraphics[width=\columnwidth]{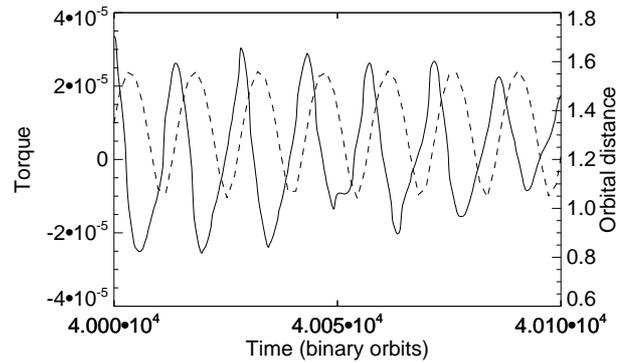}
\caption{ Evolution of the orbital radius 
(dashed line) of Kepler-34b and evolution of the specific torque 
(solid line) experienced by the planet over  $\sim 5$ orbital periods for the  run with $h=0.05$ and 
$\alpha=10^{-2}$.}
\label{k34}
\end{figure}
The middle panel of Fig. \ref{mf} shows the temporal evolution of the semi-major axis and eccentricity
of a planet with $q\sim 1.1\times 10^{-4}$, the mass ratio reported for Kepler-34b.
Eq. \ref{gapopening} suggests that a shallow gap can be formed in the calculation with $h=0.03$, 
and in the one with $h=0.05$ and $\alpha=10^{-4}$. Significant gap formation is not expected in the
other runs.

The evolution is found to be unstable in the simulation with $h=0.07$, with the planet experiencing a
close encounter with the central binary at $t\sim 1.5\times 10^4$ P. This is likely to arise because  
the disc is highly eccentric (as indicated by the left middle panel of Fig. \ref{eccentricity}), which drives a large
planet eccentricity.  

During the runs with $h=0.05$ and $\alpha \ge 10^{-3}$ the planet migrates inward until the 
eccentricity increases to $e_p\sim h$, at which point migration stalls due to torque reversal. 
A planet on a significantly eccentric orbit has smaller angular velocity at apocentre than 
neighbouring disc material, resulting in the outer disc exerting a positive torque on the planet.
The opposite happens at pericentre and the inner disc exerts a negative torque. In general
the total orbit-averaged torque experienced by a planet is expected to go from having negative to positive values
as the eccentricity grows above values $e_p > h$ (Papaloizou \& Larwood 2000). For a single planet
embedded in a disc this does not normally imply that migration changes from being inward to outward
because energy loss associated with eccentricity damping still causes the semi-major axis to
decrease, albeit more slowly than for a circular orbit (Cresswell et al 2007). When the eccentricity
is maintained through interaction with the central binary, however, then the total time averaged torque 
can equal zero and inward migration can be halted. In order to demonstrate that this effect is at work 
we follow Pierens \& Nelson (2008) and plot the disc torque and the planet orbital position over a few 
planet orbital periods for the run with $\alpha=10^{-2}$ in Fig. \ref{k34}.
As expected, the torque exerted on the planet is positive (resp. negative) when the planet is at apocentre
(resp. pericentre) and the orbit-averaged torque almost cancels. As noted by Pierens \& Nelson (2008), the 
slight phase shift between the two curves is due to the time needed for the planet to create an inner 
(outer) wake at apocentre (pericentre).
While Kepler-34b is observed to orbit at $a_p\sim 1.1$ AU from the central binary,  we note 
that the final planet semi-major axis in the run with $h=0.05$ and $\alpha=10^{-3}$ is 
$a_p\sim 1.45$ AU whereas it is $a_p\sim 1.3$ AU in the calculation with $\alpha=10^{-2}$. Interestingly, 
the final value for the eccentricity $e_p \sim 0.2 $ obtained in both runs is rather close from the observed value 
($e_p\sim 0.18$). 

\begin{figure}
\includegraphics[width=\columnwidth]{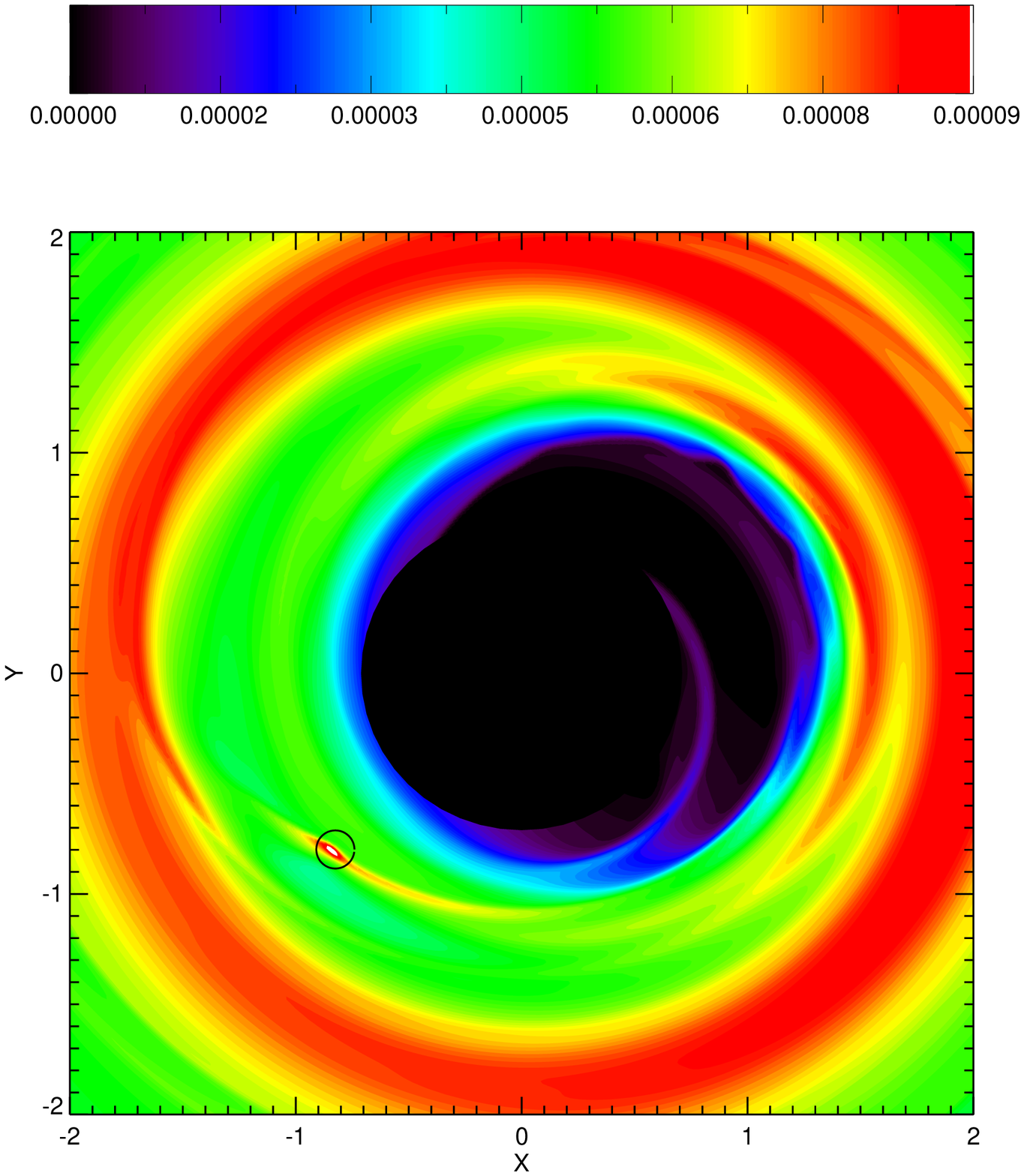}
\includegraphics[width=\columnwidth]{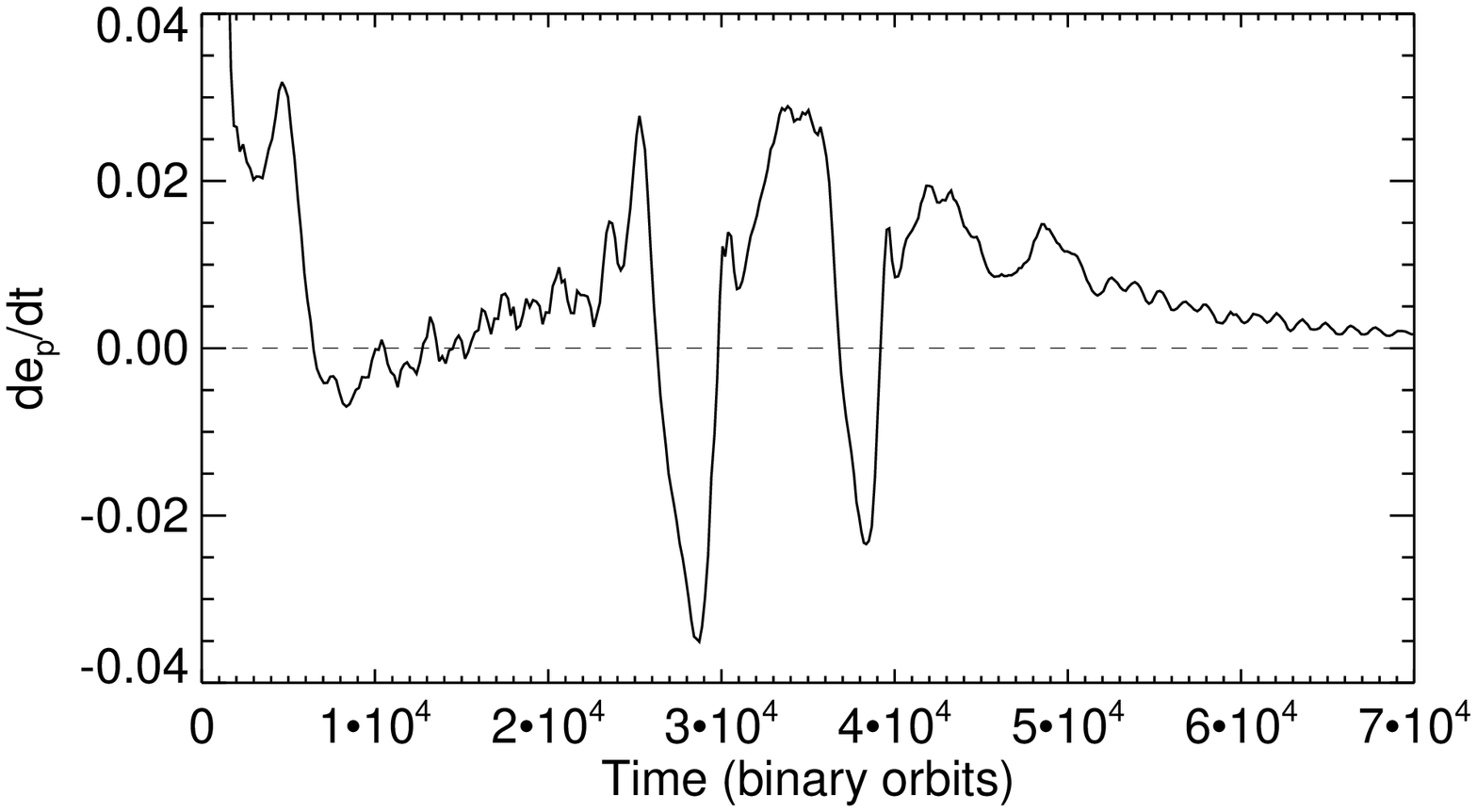}
\caption{{\it Upper panel:} Snapshot of the disc surface density at t= for the Kepler-34b run with $\alpha=10^{-4}$ and $h=0.05$. 
{\it Lower panel:} Time evolution for the change of the planet eccentricity induced by the disc.} 
\label{2d-kep34}
\end{figure}

For the two runs with ($h=0.03$, $\alpha=10^{-3}$) and ($h=0.05$, $\alpha=10^{-4}$) early evolution proceeds 
similarly, with the planet migrating inward until its eccentricity has increased to a value of $e_p\sim 0.2$, 
at which point migration stops.
As mentioned above, however, the planet can significantly perturb the disc surface density profile in these two cases 
and this causes the planet to migrate inward at later times because the balance of torques is modified.
In both cases the planet migrates inward again from $t\sim 1.5\times 10^4$ binary orbits but the final outcome
of each run differs. In the $h=0.03$ run the planet eccentricity decreases somewhat as the planet migrates deep inside the 
truncated cavity. Continuation of the run until $\sim 1.5\times 10^5$ binary orbits shows that migration eventually
halts when the total orbit-averaged torque vanishes due to the finite eccentricity (see discussion above).  
At the end of the run the planet semi-major axis reaches a constant value $a_p\sim 1.35$ AU and the eccentricity $e_p\sim 0.1$, 
much smaller than the observed eccentricity of Kepler-34b. 
In the simulation with $h=0.05$ and $\alpha=10^{-4}$ significant growth of the planet eccentricity occurs at 
$t\sim 2.5\times 10^4$ binary orbits, causing the planet to undergo an episode of outward migration.
Analysis of this simulation shows that this sudden growth of $e_p$ is caused by interaction between the planet and 
an arc-shaped high density structure related to the $m=1$ eccentric mode in the disc, 
with pattern speed of $\sim 2\times 10^3$ binary orbits. 
This is illustrated in the upper panel of Fig. \ref{2d-kep34} which shows a snapshot of the disc surface density at 
$t\sim 2.5\times 10^4$ binary orbits. In the lower panel of Fig. \ref{2d-kep34}, we plot the temporal evolution of 
the theoretical change of the planet eccentricity $\dot e_p$ due to the disc with (e.g. Bitsch \& Kley 2010):    
\begin{equation}
\frac{\dot e_p}{e_p}=\frac{1-e_p^2}{e_p^2}\left(\frac{1}{2}\frac{\dot a}{a}-\frac{T_d}{J_p}\right)
\end{equation}
where $\dot a$ is the estimated planet orbital decay rate, $T_d$ is the torque exerted by the disc 
on the planet and $J_p$ is the angular momentum of the planet. This figure clearly demonstrates that the disc 
is responsible for driving the planet eccentricity upward at $t\sim 2.5\times 10^4$. At later times it appears 
that the combined effect of eccentricity growth and outward migration cause the planet 
to be in contact with the disc density bump located at $\sim 1.8$ AU (see upper panel of Fig. \ref{2d-kep34}), corresponding
to material with higher specific angular momentum than the planet. Strong damping of the planet eccentricity occurs
at $t\sim 2.7\times 10^4$ and the planet migrates inward again. The long-term trend is of decreasing
semi-major axis and eccentricity until the torque on the planet becomes negligible.  
At $t\sim 6\times 10^4$ binary orbits the planet semi-major axis is $a_p\sim 1 $ AU and its 
eccentricity has reached a quasi-stationary value of $e_p\sim 0.03$, much smaller than the observed 
value ($e_p\sim 0.18$). The low value for $e_p$ arises because the planet orbits deep inside the tidally-truncated
cavity where interaction with the eccentric disc is small. It is also related to the fact that $q_b\sim 1$ 
for Kepler-34, so that the forced eccentricity due to interaction with the binary is small. Continuation 
of this run for $\sim 1.5\times 10^5$ binary orbits confirms that the planet migrates very slowly into the 
gas depleted cavity on a near-circular orbit, but too close to the central binary to be consistent 
with the observed orbital parameters of Kepler-34b. At $t\sim 1.5\times 10^5$ binary orbits, the planet semi-major axis 
is $a_p\sim 0.95$ AU but there remains evidence that inward migration has not stalled completely.

\subsection{Kepler-35b}
The orbital evolution of the Kepler-35b simulations are shown for the different disc models in the lower panel of 
Fig. \ref{mf}. Compared with Kepler-34b, for which $q\sim 1.1\times 10^4$, the mass ratio of Kepler-35b is 
$q\sim 7.5\times 10^{-5}$. Despite the lower binary eccentricity the results for this system look rather similar 
to the corresponding Kepler-34b runs. 

For disc models with $h\ge 0.05$ and $\alpha\ge 10^{-3}$ the evolution outcomes are 
stalling of inward migration leaving the planet on an eccentric orbit at fixed semi-major axis. 
As we have seen in the previous section, this quasi-equilibrium state is reached when the large 
positive torque experienced by the planet at apocentre is in exact balance with the weaker negative 
torque felt by the planet at pericentre. The final position of the planet is clearly correlated with 
the size of the tidally truncated inner cavity. For example, in the run with $h=0.07$ and 
$\alpha=10^{-3}$, the density peak is located at $R\sim 1.3$ AU (see lower right panel of Fig. \ref{eccentricity}) 
and the final planet semi-major axis is $a_p\sim 1.25$ AU whereas in the simulation with $h=0.05$ and $\alpha=10^{-3}$, 
the density peak is at $R\sim 1$ AU and migration stops at $a_p\sim 0.9$ AU. Not surprisingly, the final value for 
the planet eccentricity appears to scale with the disc eccentricity. For instance, in the run with  
$h=0.05$ and $\alpha=10^{-3}$, the disc eccentricity is $e_d\sim 0.05$ (see left panel of Fig. \ref{eccentricity}) 
and the final $e_p\sim 0.11$. In the run with $\alpha=10^{-2}$, the disc eccentricity $e_d\sim 0.07$ and 
$e_p\sim 0.21$ at the end of the calculation.

A similar outcome is observed in the run with $h=0.03$. Here the planet migrates rapidly inward 
at early times until it reaches a stable, non-migrating configuration with $a_p\sim 0.7$ AU and 
$e_p\sim 0.06$. Although not shown here, analysis of the torque experienced by the planet 
clearly reveals a positive (resp. negative) disc torque at planet apocentre (resp. pericentre) 
and a zero time-averaged total torque. 

For the simulation with $h=0.05$ and $\alpha=10^{-4}$ the process described above makes migration stall
at $t\sim 10^4$ binary orbits once the planet eccentricity has reached $e_p\sim 0.12$ and its semi-major 
axis $a_p\sim 1.1$ AU. Similarly to the case of Kepler-34b, the planet is found to  migrate inward again at 
later times due to gap formation by the planet. Subsequent evolution again corresponds to the planet migrating deep 
inside the inner cavity until the local disc mass becomes too small for further substantial migration.   
Despite a runtime of $\sim 1.5\times 10^5$ binary orbits, an equilibrium state was not reached for this 
case. At the end of the run, the planet semi-major axis is $a_p\sim 0.69$ but inward migration still 
proceeds. Its eccentricity, however, has reached a constant value of $e_p\sim 0.01$. Again, such a low value for 
$e_p$ with respect to the observed one ($e_p\sim 0.04$) arises because interaction betwen the planet and the 
eccentric disc is weak inside the cavity, and the binary mass ratio $q_b\sim 1$ for Kepler-35 so the forced
eccentricity remains small.

\section{Evolution of migrating and accreting protoplanets}
We now consider a scenario in which a $\sim 20$ M$_{\oplus}$ core forms in the disc at large 
distance from the central binary and migrates
inward {\it via} type I migration until it reaches a location where migration halts.  For a 
$20$ M$_{\oplus}$ planet and 
for the fiducial value of the disc aspect ratio $h=0.05$, the 
half-width of the horseshoe region is $x_s\sim 1.1a_p\sqrt{q/h}\sim 0.027 a_p $ (Paardekooper et al. 2010), which means 
that at $1$ AU ($R=2$ in the computational domain), the horseshoe region is resolved by about $5$ grid 
cells in the radial direction. For such a resolution, the relative error on the corotation torque is estimated 
to be $\sim 15\%$ (Masset 2002). Once migration is stalled, the planet then accretes gas
from the disc until the disc has been dispersed by photoevaporation. The primary reason for including photoevaporation
is that the planets in the Kepler-16, Kepler-34 and Kepler-35 systems are $\sim$ Saturnian mass, and we
know from calculations of giant planet formation that such planets are able to accrete gas at very rapid
rates (e.g. Pollack et al 1996). It seems likely that Saturn-mass planets embedded in a substantial reservoir 
of gas will grow rapidly, so we work with the assumption that growth beyond their observed masses was prevented
by removal of the gas disc.  

Our procedure is as follows. We first calculate the orbital evolution of protoplanets with mass 
ratio $q=3\times 10^{-5}$ interacting with the different circumbinary disc models to illustrate
the behaviour for different disc parameters. A typical outcome is stopping of inward migration close to the edge 
of the tidally truncated cavity, and this arises when the planet eccentricity reaches a value $e_p\sim h$. For 
each of the Kepler binary systems we restart the simulation which uses the disc model that best reproduces the orbital 
parameters of the observed planet among the simulations presented in Sect.~4, switching on gas accretion onto the core. 
The final mass and orbital parameters of the planet are set by the gas dispersion timescale.

\subsection{Evolution of a protoplanetary core}
\label{sec:core}
\begin{figure*}
\includegraphics[width=\textwidth]{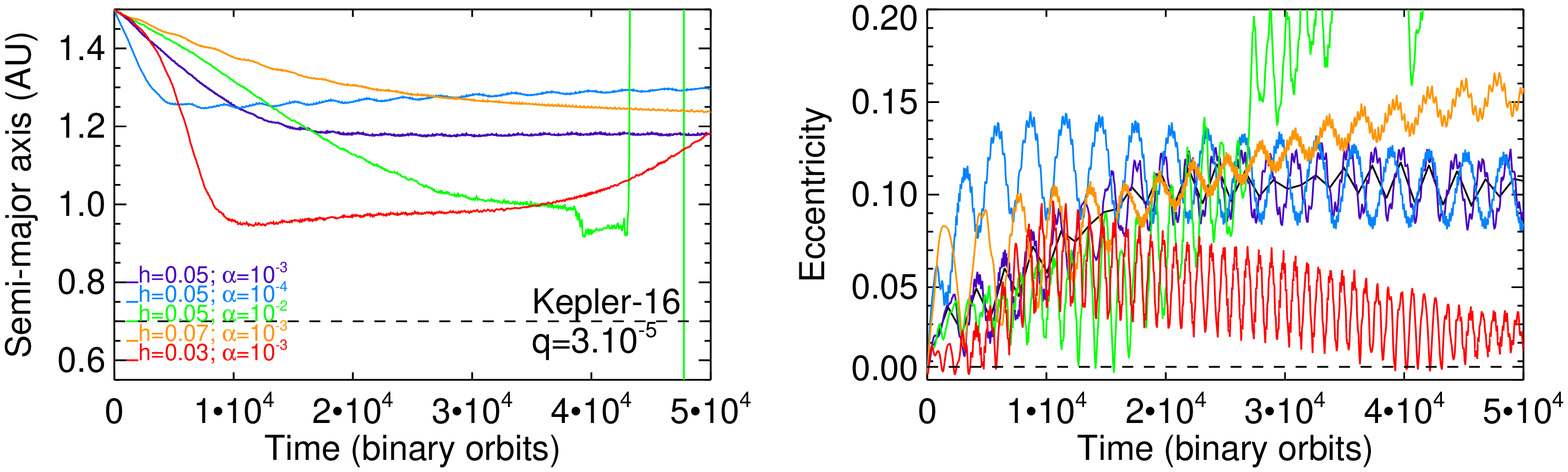}
\includegraphics[width=\textwidth]{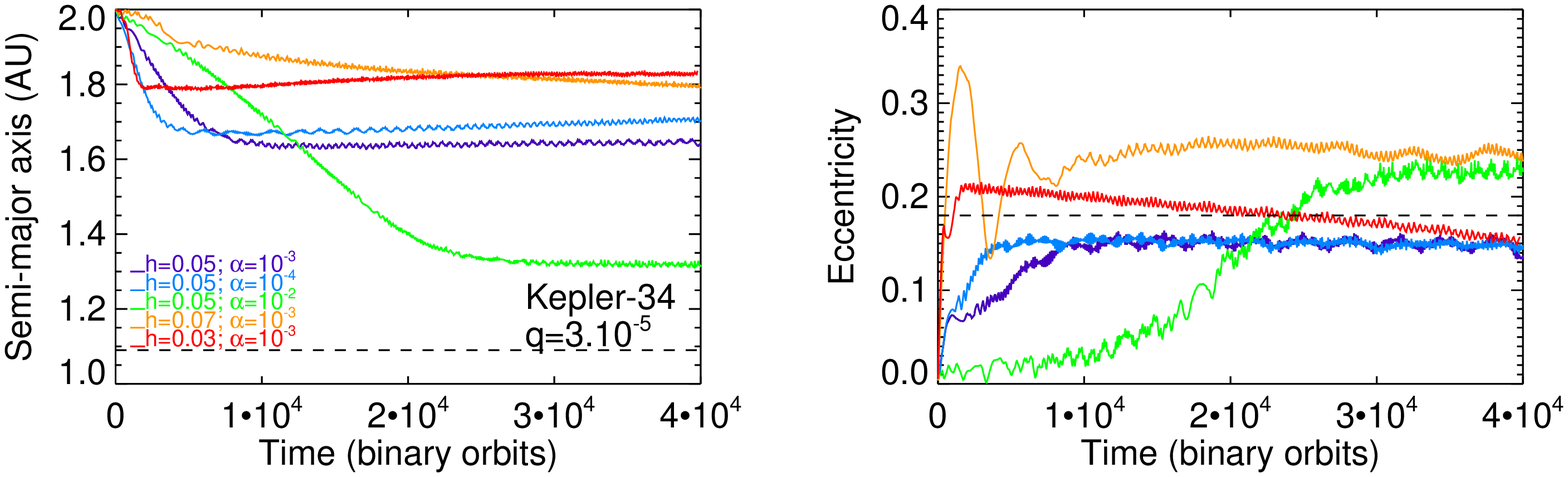}
\includegraphics[width=\textwidth]{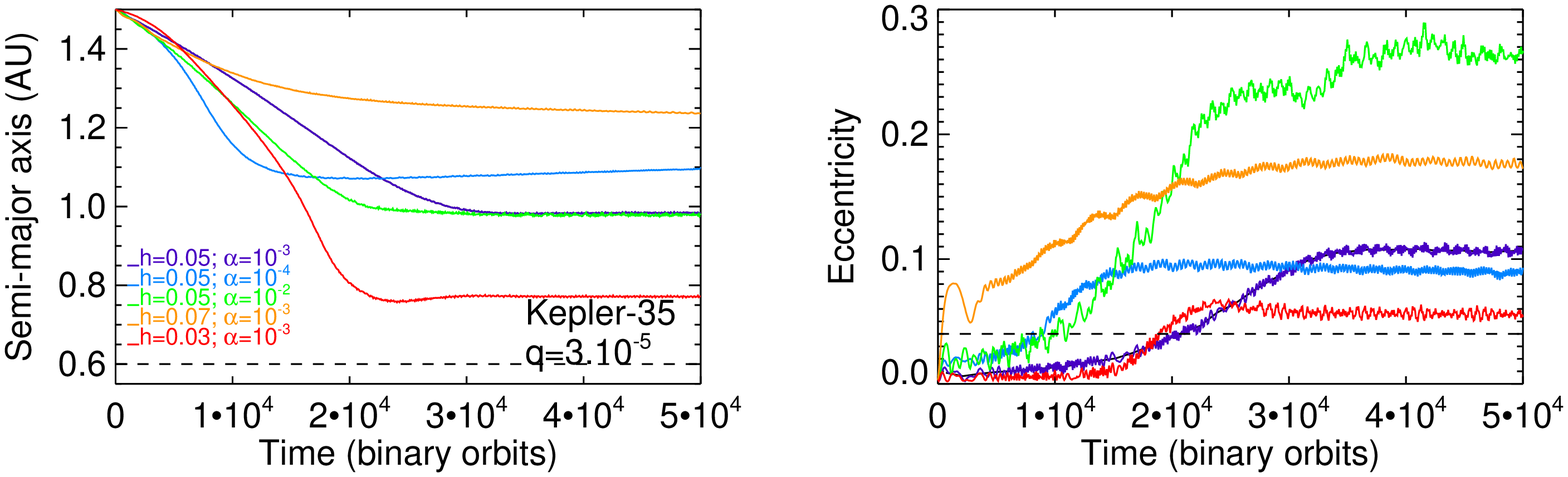}
\caption{{\it Left panel:} Time evolution of the semi-major axis of an embedded protoplanet 
with mass ratio $q=3\times 10^{-5}$ . {\it Right panel:} Time 
evolution of the planet eccentricity. For each parameter, the dashed line correspond
to the observed value} 
\label{20embryo}
\end{figure*}
For the different isothermal disc models we consider, the orbital evolution of an embedded circumbinary protoplanet with mass ratio 
$q= 3\times 10^{-5}$ is presented in Fig. \ref{20embryo}. The upper panel shows the results of 
simulations in which the binary parameters are those of Kepler-16. The middle and lower panels correspond to the Kepler-34 and Kepler-35 systems. 

\begin{figure}
\includegraphics[width=\columnwidth]{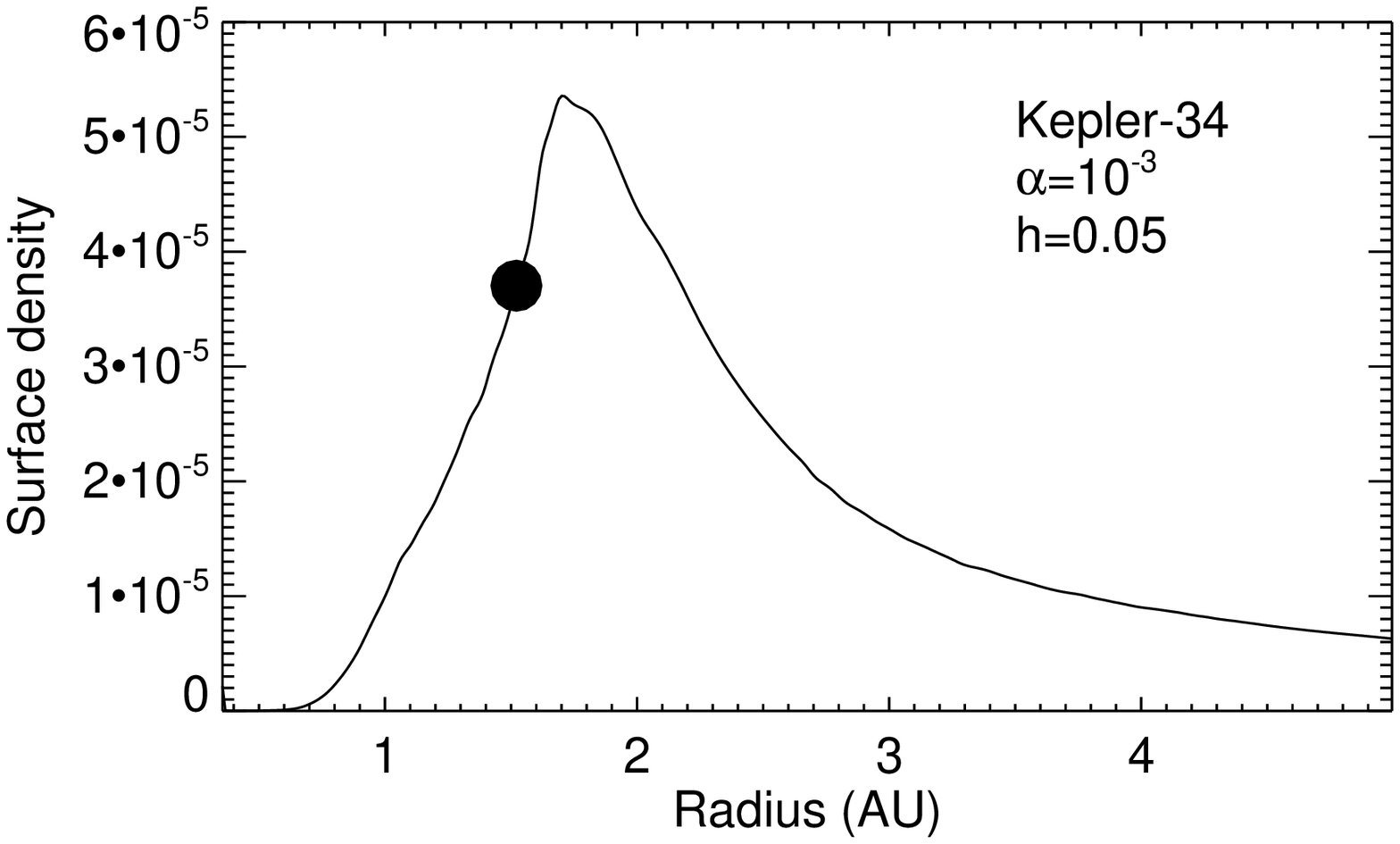}
\includegraphics[width=\columnwidth]{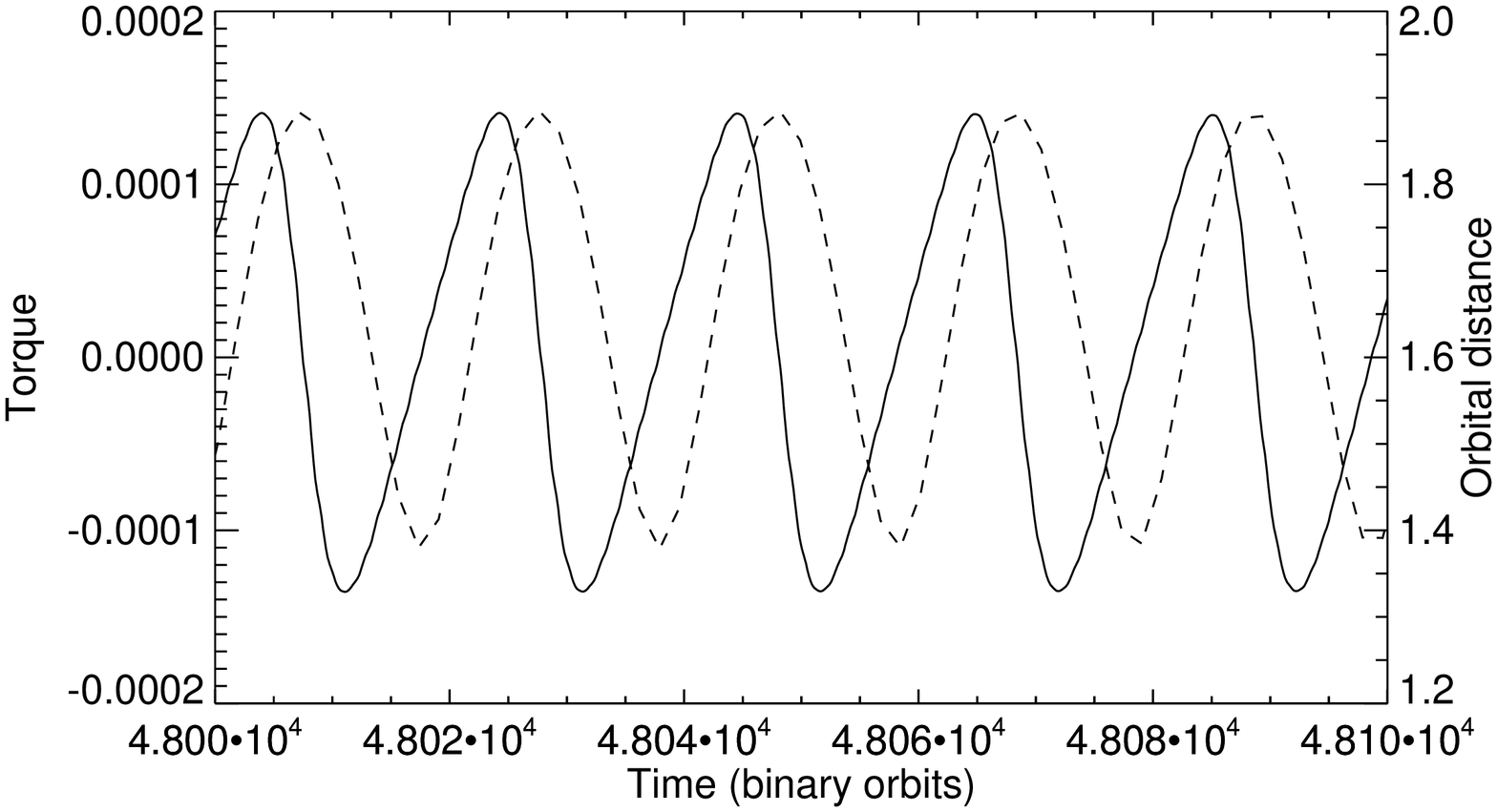}
\caption{{\it Upper panel:}   Surface density profile and planet position for the  Kepler-34 
run with $h=0.05$ and $\alpha=10^{-3}$ at $t=2\times 10^4$ binary orbits. {\it Lower panel: } Evolution of the orbital radius 
(dashed line) of a planet with mass ratio $q=3\times 10^{-5}$ and evolution of the specific torque 
(solid line) experienced by the protoplanet over  $\sim 5$ orbital periods for the same run.}
\label{denm20}
\end{figure}

We begin presentation of the simulation results by focusing on these two latter cases, for which similar outcomes are
observed. Early evolution typically involves inward migration of the protoplanet and continued growth of its eccentricity
due to its interaction with both the central binary and the eccentric disc. At later times inward migration slows down significantly in the run 
with $h=0.07$, and completely stalls or even reverses for thinner disc models.
In these cases the final location of the planet is clearly correlated with the position of the gap edge in the right panel of 
Fig. \ref{eccentricity}. For example, considering the Kepler-34 system, 
the density peak is located at $R\sim 2$ AU in the simulation with $h=0.03$ while the middle panel of 
Fig. \ref{20embryo} shows that migration stops at $R\sim 1.8$ AU for the same disc model. 
In the fiducial run with $h=0.05$ and $\alpha=10^{-3}$, the density peak is located at $R\sim 1.5$ AU and the planet migrates to
$a_p\sim 1.5$ AU before migration halts. We plot  both  the surface density profile and the planet position at $t=2\times 10^4$ 
binary orbits in the upper panel of Fig. \ref{denm20} for this run, showing that the planet halts at the edge of the cavity.
As with a number of the simulations described in Sect.~3, the halting of migration occurs because of the torque
reversal that occurs when $e_p \ge h$. For eccentric orbits the corotation torque is expected to be substantially
diminished (e.g. Bitsch \& Kley 2010), so they cannot be responsible for the halting of migration.
Following a  strategy  similar to that described in Sect. \ref{sec:kepler34}, where the correlation 
between the disc torque and the planet orbital radius is examined, we confirm that at late times the planet experiences a positive torque at apocentre and a negative 
torque at pericentre, leading to a null time-averaged total torque. This is illustrated in the lower panel of
Fig. \ref{denm20} where we plot both the disc torque and the planet orbital position over a few orbital periods of the planet 
for the run with $h=0.05$ and $\alpha=10^{-3}$. 

The outward migration that we observe in simulations with $h\le 0.05$ and $\alpha\le 10^{-3}$ is likely related to the steepness of the inner cavity edge 
for these disc models, which combined with the fact that the planet spends more time at apocentre, favours a net positive torque on the planet. 
In runs with $h=0.07$, however, the gradient of surface density in the inner cavity is shallower and so less favourable for stalling
or reversing of migration for an eccentric planet. This explains why type I migration is not halted in that case.

The results of the Kepler-16 runs are displayed in the upper panel of Fig. \ref{20embryo}. Overall, these are consistent with the corresponding 
Kepler-34 and Kepler-35 simulations, except for the runs with ($h=0.03$, $\alpha=10^{-3}$) and ($h=0.05$, $\alpha=10^{-2}$) in which the final fate 
of the planet is found to be very different. In the simulation with $h=0.03$, migration reverses at $\sim 10^4$ binary orbits due to the eccentric
planet orbit, but outward migration accelerates from $t\sim 3.5\times 10^4$ binary orbits onward. Inspection of the apsidal lines for both the disc 
and the planet reveals that from this time these become misaligned, with a tendency for the apsidal line of the disc to slightly lead that of the planet. 
This promotes outward migration because the planet interacts preferentially with matter whose angular velocity is greater than that of the planet at 
apocentre (Pierens \& Nelson 2008). As outward migration proceeds, the planet eccentricity progressively decreases. 
Continuation of the run shows that the planet eventually migrates inward again once its semi-major axis has reached $a_p\sim 1.3$ AU. 
Therefore we can reasonably expect the long term evolution for this run to consist of periods of inward followed by outward migration until disc dispersal.

With respect to the other disc models, the positive gradient of  surface density at the inner edge of the 
tidally truncated cavity is shallower in the simulation with  ($h=0.05$, $\alpha=10^{-2}$) (see upper right panel of Fig. \ref{eccentricity}). 
Consequently, torque reversal is not observed in that case and the planet migrates inward continuously. This inward migration is accompanied 
by significant growth of the planet eccentricity due to its interaction with both the binary and eccentric disc.  The eccentricity increases to
$e_p\sim 0.3$ at $t\sim 4\times 10^4$ P. At $t\sim 4.3\times 10^4$ $P$, the eccentricity growth causes the planet to undergo a close encounter with 
the binary and to be subsequently ejected from the system.    

\subsection{Evolution of an accreting core}
In this section we use the disc model that best reproduces the observed parameters of each of the three 
Kepler systems to focus on a more realistic evolution scenario in which a migrating embryo grows by 
accreting gas from the disc at the same time as the disc is being dispersed. We expect both core growth and
inward migration to be terminated as a result of disc clearing, which is thought to be mainly driven by 
photoevaporation induced by UV irradiation and/or X-rays at late times (Owen et al. 2011).
Gas disc dispersal by photoevaporation is modelled in the simplest possible manner by forcing the gas surface density 
to decay exponentially with an e-folding time $t_{dis}$, for which we consider values $10^4$ and $10^5$ P. 
These values are broadly consistent with estimates of the gas dispersal timescale due to UV photoevaporation which 
give $t_{dis}\sim 10^5$ years (Alexander et al. 2006). We also vary the rate at which the planet accretes gas from 
the disc by choosing different values for the accretion parameter, $f_{acc}$, defined in Sect.~\ref{sec:init}. 
For each value of disc dispersal time $t_{dis}$ we perform simulations for values of $f_{acc}=1$, 2, 3, ..., 10. 
We note that $f_{acc}=1$ corresponds to the maximally accreting planet (e.g. Kley 1999).
\subsubsection{Kepler-16}
A core mass of $m_p\sim 20\; M_\oplus$ is close to the value required for runaway gas accretion in
the core accretion scenario of gas giant planet formation (Pollack et al. 1996; Papaloizou \& Nelson 2005),
which is why this core mass was chosen for these simulations. The disc model that gave best agreement
with the observed orbital parameters of Kepler-16b in Sect.~4.1 had $h=0.05$ and $\alpha=10^{-4}$, so
we consider a model that was run with planetary mass ratio $q=6.7\times 10^{-5}$ ($m_p\sim 20\; M_\oplus$)
until migration halts, and then restart with gas accretion and photoevaporation switched on for the model
parameters described above. 

The evolution of the model that best fits the observed Kepler-16b system is shown in the top panels of 
Fig. \ref{accretion}, and had parameters $f_{acc}=1$ and $t_{dis}=10^5$ P. The left panel shows the planet 
and disc masses, and the right panel shows the semi-major axis and eccentricity. The planet mass
slightly exceeds the inferred mass of Kepler-16b, the semi-major axis stalls just outside of the
observed value, and the eccentricity has a mean value $e_p \simeq 0.03$ (to be compared with the
smaller value of $7 \times 10^{-3}$ inferred from observations). It seems very likely that
subtle tweaking of the model parameters could lead to a planet whose mass and semi-major axis
agrees much more accurately with the observed values, but our aim here is not to obtain
perfect agreement with the observed values using a highly-simplified disc model so
we do not attempt to refine the values of $f_{acc}$ and $t_{dis}$ here. Obtaining a mean eccentricity 
equal to the very small value reported for this system would appear to present a greater challenge.
It remains to be seen whether a more realistic disc model will lead to a smaller final planet eccentricity.

\subsubsection{Kepler-34}
The simulations in Sect. 4.2 indicate that the disc parameters $h=0.05$ and $\alpha=10^{-2}$
best produce a final planetary configuration that approximates the observed values. The
simulation with planetary mass ratio $q=3 \times 10^{-5}$ described in Sect.~5.1 above was 
restarted with gas accretion and disc dispersal switched on, and the results from the
run whose outcome best matches the observed Kepler-34b system are shown in the
middle panels of Fig. \ref{accretion}. This simulation adopted $f_{acc}=3$ and $t_{dis}=10^4$ P.
We see that the simulated planet mass undershoots the observed value slightly (a slightly
larger value of $t_{dis}$ would presumably improve this). We see that the semi-major
axes and eccentricities barely change at all during the time when gas accretion occurs,
suggesting that the final orbital configuration obtained for the 20 M$_{\oplus}$ embryo is 
very similar to that for the final planet after accretion. Inspection of Fig.~\ref{mf}
shows that migration of the fully-formed planet also leads to very similar
final values for $a_p$ and $e_p$. Unlike in the Kepler-16 run, the simulations of
Kepler-34b using the disc model $h=0.05$ and $\alpha=10^{-2}$
produce the same set of orbital elements when considering a fully-formed planet, a
migrating 20 M$_{\oplus}$ core, and a gas-accreting core. Producing a closer approximation
to the observed system will therefore require modest tweaking of the disc physics
rather than the parameters $f_{acc}$ and $t_{dis}$.

\subsubsection{Kepler-35}
\begin{figure*}
\includegraphics[width=\textwidth]{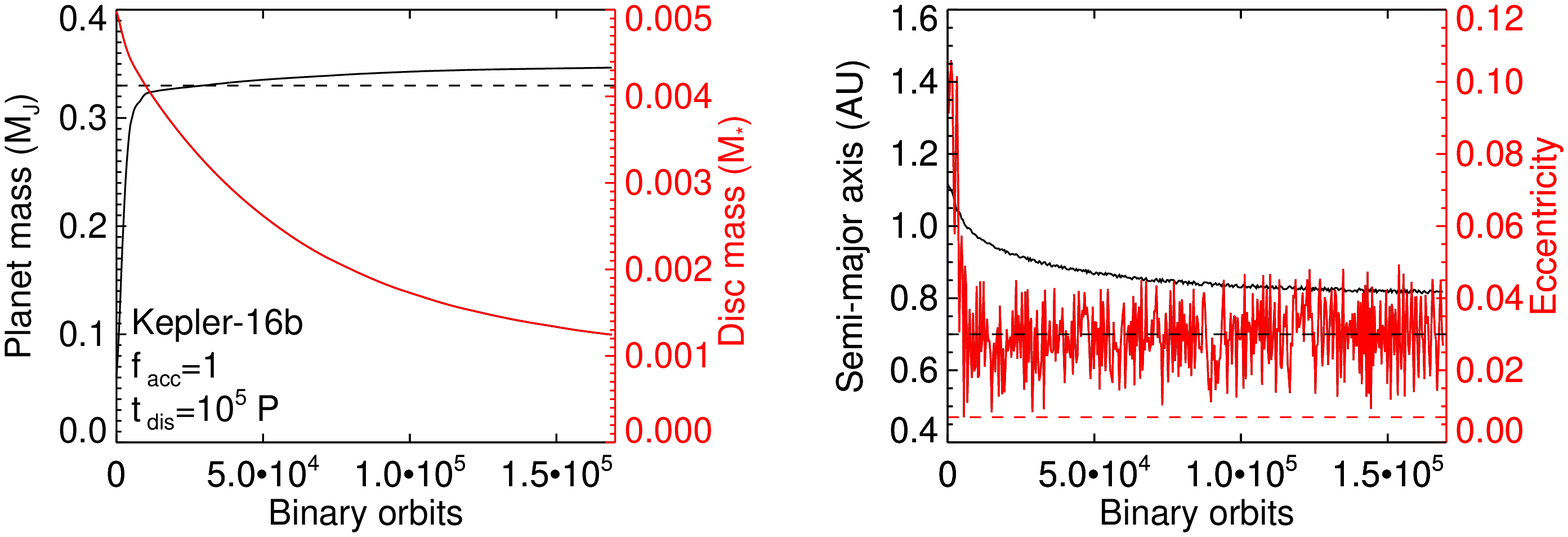}
\includegraphics[width=\textwidth]{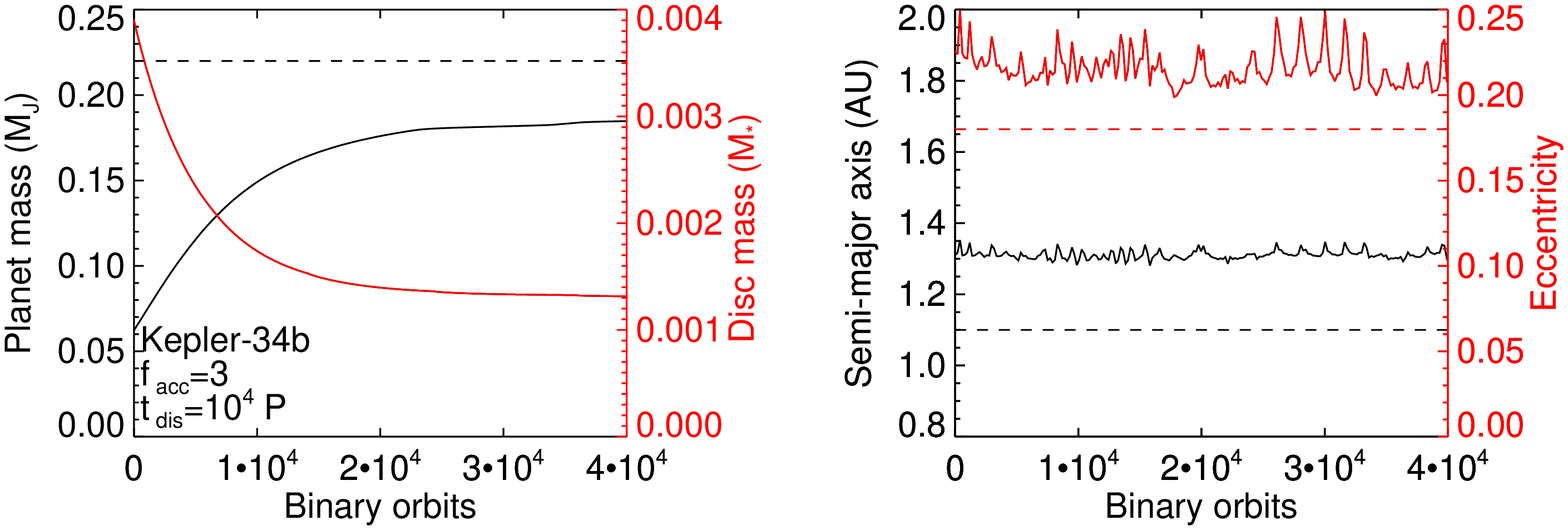}
\includegraphics[width=\textwidth]{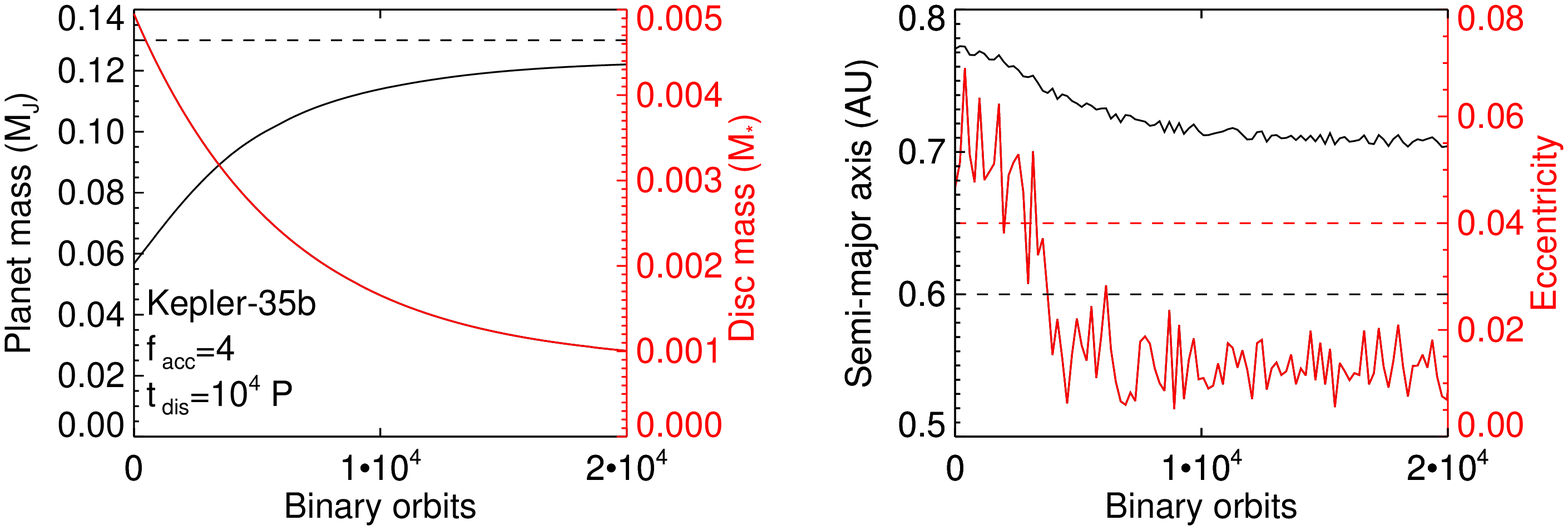}
\caption{{\it Left panel:} Time evolution of the planet mass (black) and disc mass (red) for simulations in which a 
migrating embryo grows by accreting gas material from the disc. {\it Right panel:} Time 
evolution of semi-major axis (black) and eccentricity (red). For each parameter, the dashed line correspond 
to the observed value.}
\label{accretion}
\end{figure*}

The disc model with $h=0.03$ and $\alpha=10^{-3}$ was shown to be the best candidate for
reproducing the orbital configuration of Kepler-35b in Sect.~4.3. 
The simulation with planetary mass ratio $q=3 \times 10^{-5}$ described in Sect.~5.1 above was
restarted with gas accretion and disc dispersal switched on, and the results from the
run whose outcome best matches the observed Kepler-35b system are shown in the
lower panels of Fig. \ref{accretion}. This simulation adopted $f_{acc}=4$ and $t_{dis}=10^4$ P,
and produces a planet mass close to the one observed, but a semi-major axis
that is larger than the observed value ($a_p=0.7$ AU versus 0.6 AU). This is consistent with
the corresponding run with the fully formed planet in Sect.~4.3 which also resulted in a
planet with $a_p \simeq 0.7$. Reducing this value to agree with the observed one will
therefore require moderate adjustment of the disc model. The eccentricity evolution
results in a final mean value of $e_p \simeq 0.015$. Interestingly this is smaller than
the observed value of 0.04, unlike the value obtained from the simulation of the
fully-formed planet which resulted in $e_p \simeq 0.06$ (see Fig.~\ref{mf}).

 It is clear from this and the
other simulations described for Kepler-35b and Kepler-16b that obtaining fits to the
observations for all system parameters would require implementation of a multi-dimensional fitting
procedure whose complexity would not be justified given the simplicity of our underlying model.

\subsection{Validity of the isothermal approximation}
\label{sec:valid}
In the locally isothermal limit, the only contribution to the corotation torque is the vorticity-related 
corotation torque which scales with the vortensity (i.e. the ratio between vorticity and 
surface density) gradient. In the context of circumbinary discs, it has been shown that at the inner edge of the tidally truncated cavity where the surface density has
a large positive gradient, the  (positive) vorticity-related corotation torque can be strong enough to 
counterbalance the (negative) differential Lindblad torque, leading to halting of Type I migration 
(Masset et al. 2006; Pierens \& Nelson 2007). This arises provided that the amplitude of the corotation torque is 
close to its optimal value, which is reached when:  i) the viscous diffusion 
timescales across the horseshoe region are approximately equal to half the libration timescale (Baruteau \& Masset 2013);  ii) the planet eccentricity 
$e_p$ is significantly smaller than the dimensionless half-width of the horseshoe region $\hat x_s$. We expect indeed that when  $e_p\sim \hat x_s$, the radial 
excursion of the planet becomes larger than the half-width of the horseshoe region and consequently the corotation 
torque to be strongly attenuated (Bitsch \& Kley 2010).
For  $e_p\sim 0.1$, which is a value typical to what is observed in the isothermal simulations of $20\;M_\oplus$ cores presented 
above,  
it is likely that the contribution of the corotation torque to the total torque is relatively weak 
(see Fig. 2 of Bitsch \& Kley 2010).\\ 
In non-isothermal discs, the corotation torque consists of the vorticity-related corotation torque 
plus an additional component which scales with the entropy gradient (Paardekooper \& Papaloizou 2008; 
Baruteau \& Masset 2008).  For a negative entropy gradient inside the disc, this entropy-related corotation 
torque is positive and can be responsible for significantly slowing down or even reverse Type I migration. Depending on 
whether or not a locally isothermal equation of state is used and provided that the amplitude of the corotation 
torque is close to its fully-unsaturated value,  simulations of the evolution of  $20\;M_\oplus$ cores in the Kepler-16, Kepler-34 and Kepler-35 systems  may therefore lead to very different outcomes.  Regarding the simulations with fully-formed 
planets, we note that the issue of the equation of state does not really matter since we expect a Saturn-mass planet
to open a partial gap in the disc, with the consequence that the corotation torque is weakened (Kley \& Crida 2008).\\
Assuming a $20\;M_\oplus$ core forming at $\gtrsim 10 $ AU (Marzari et al. 2013), namely in a region of small 
disc eccentricity, its migration can 
be directed 
either outward or inward in the non-isothermal case,  depending on the level of saturation of the corotation torque. In both cases, however, we expect 
the protoplanet to reach a zero-migration line where it becomes trapped (Lyra et al. 2010; Hellary \& Nelson 2012). As the disc disperses, 
the zero-migration line is shifted inward with the main consequence that  the planet eventually enters a region of high disc eccentricity. In that case, the 
embryo should decouple from the zero-migration line at a point in time where its eccentricity becomes high enough 
for the corotation torque to be significantly weakened. The planet then migrates inward again and 
we expect evolution outcomes rather similar to those arising in isothermal runs.  This implies that in the phase 
after the embryos eccentricities have reached values $e_p>\hat x_s$,  the results of 
isothermal runs presented above are relatively robust regarding the equation of state.\\
 In disc models with moderate values 
of disc eccentricity, if any, an alternative outcome is that the planet moves inward as the disc disperses until it reaches 
the inner edge of the inner cavity where it becomes trapped due to the action of the vorticity-related corotation 
torque. \\
As the planet eccentricity is mainly driven by the eccentric disc, this raises the issue of the disc eccentricity in the radiative case. Clearly, this issue needs to be examined in more details 
and we will address this issue and its consequence on the evolution of protoplanetary cores in the Kepler-16, 
Kepler-34 and Kepler-35 systems in a future publication.
\section{Discussion and conclusion}
In this paper, we have presented the results of hydrodynamic simulations that examine the 
migration and orbital evolution of embedded circumbinary planets orbiting around the three 
close binary systems Kepler-16, Kepler-34 and Kepler-35. We began our study by examining
the evolution of the binary-plus-disc system in the absence of an embedded planet in order
to obtain an improved understanding of how circumbinary discs are structured by the central binary
as a function of binary and disc parameters. In particular, we examined how the mean surface
density radial profile and disc eccentricity vary for the three Kepler binary systems as
a function of disc aspect ratio, $h$, and viscosity parameter $\alpha$. Our results 
indicate that for each binary system the disc structure is a sensitive function of the
disc parameters, with the tidally trucated cavity displaying significant changes in both
radius and internal density structure as a function of $h$ and $\alpha$. Furthermore,
the eccentricity of the disc also shows strong dependence on disc parameters: we observe
a general tendency for the disc eccentricity to increase with increasing $h$, which is likely
due to the fact that the eccentricity is more effectively communicated through the disc 
for larger values of $h$. 

Our investigation confirms that the origin of the disc eccentricity is non linear mode coupling 
as described previously for circumbinary and circumstellar discs by Papaloizou et. al. (2001) and 
Lubow (1991), respectively. The forced eccentricity due to secular interaction with the
binary is too small to explain the simulations (see Sect.~3.1). The Papaloizou et al (2001) study indicated 
that coupling between an initial $m=1$ eccentric disc disturbance and the $m=1$ component of the
binary potential with pattern speed equal to the mean motion of the binary
leads to excitation of a wave at the 3:1 eccentric Lindblad resonance which
transports angular momentum outward, increasing the disc eccentricity.
Our study indicates that the 3:1 resonance is important also for the 
eccentric Kepler binary systems, but also suggests that mode coupling
involving higher order terms in the binary potential may also be important.
Given the importance of the disc eccentricity for determining the orbital
parameters of embedded circumbinary planets, a further, more in-depth study of circumbinary
disc dynamics is certainly warranted.

Our study of the evolution of circumbinary planets embedded in the circumbinary discs
described above adopted two basic scenarios. In the first we considered the orbital
evolution of a fully-formed planet in a variety of disc models for each binary
system, and attempted to determine which of the disc models resulted in
planet orbits that best agree with the Kepler observations. The basic mode of
evolution in these runs was inward migration of the planet, followed by stalling
of migration in the tidally truncated disc cavity. The planetary orbits typically
become noncircular due to interaction with the circumbinary disc and binary, with the
former effect usually being dominant. In the second, more realistic scenario, 
we consider the evolution of a low-mass ($\sim 20$ M$_{\oplus}$) core, which
migrates inward until it stalls when it enters the inner disc cavity. Gas accretion onto
this core is then initiated, along with dispersal of the gas disc (mimicking photoevaporation
for example), until the planet achieves its final mass and orbital configuration.
We examine the outcome of this scenario as a function of gas accretion rate onto the
planet and disc dispersal time scale, and for each of the Kepler binary systems
we form circumbinary planets which are similar to the observed systems. Furthermore,
we generally find that for each of the Kepler binary systems the final orbital configuration of the planet
is similar, independent of the two migration scenarios adopted (fully-formed planet
versus migrating and accreting core).

We find that the simulation that produces the closest approximation to the observed system for
each of the Kepler binaries uses a different disc model: no single disc model
produces a final planet that is in close agreement with the observational data for all planets.
For Kepler-16b, a disc model with $h=0.05$ and $\alpha=10^{-4}$ produces the best-fit.
Good agreement with the observed planet mass and semi-major axis can be obtained, but
all models consistently predict a larger eccentricity for the planet than observed
(typically $e_p \gtrsim 0.03$ is obtained, compared with the observed value of $7 \times 10^{-3}$). 
Given that the eccentric disc is largely responsible for the eccentricity growth, this indicates
that a model with smaller disc eccentricity may be required to provide a better fit.
For Kepler-34b the disc with $h=0.05$ and $\alpha=10^{-2}$ best reproduces the orbital parameters 
of the planet. For this system a plausible model of core formation at large radius followed
by inward migration and gas accretion can be constructed that provides a good fit to the
planet mass and eccentricity, but none of our simulations were successful in forming
a planet whose final semi-major axis is as small as observed (values of $a_p \simeq 1.3$ AU
were obtained, to be compared with the observed value of 1.1 AU).
The stalling of migration occurs through an interplay between the planetary eccentricity
and the surface density profile in the inner cavity, since this is what determines the
averaged torque balance. Obtaining the required surface density profile very likely requires
a more sophisticated disc model in which either $h$ or $\alpha$ are not constant.
For Kepler-35b a disc model with $h=0.03$ and $\alpha=10^{-3}$ produces the best fit
to the planet orbital parameters. A simple model of core migration followed by gas accretion
and disc dispersal can form a planet with mass close to the observed value, but in
general the semi-major axis stalls at a value that is too large (values of $a_p \simeq 0.7$ AU
are obtained, compared with the observed value 0.6 AU). The best-fit migration-plus-accretion
model for this system produces an eccentricity that is too small (in stark contrast to the Kepler-16b runs),
with a mean value of $e_p \simeq 0.01$ being obtained, versus the observed value of 0.04.
Interestingly, the best-fit model for a fully-formed planet that migrates in and stalls
produces a value $e_p \simeq 0.06$, so the observed value falls neatly between these two 
simulated values. Again, it is likely that the details of the disc eccentricity distribution
and the surface density profile in the inner disc cavity determine the fine details of the
planetary orbital parameters, and obtaining improved fits may simply require a more
sophisticated disc model than we have considered.

In summary, our simulation results indicate strongly that the Kepler-16, Kepler-34 and Kepler-35
planets can be explained by a scenario in which a massive rocky/icy core forms at large
distance from the central binary in a circumbinary disc and migrates inward until it stalls
in the tidally truncated cavity formed by the central binary. Subsequent gas accretion leads to
the growth of the planet, and the changing balance of torques leads to a further episode of
modest inward migration. We argue that the final configurations of these planets are likely to
be determined in part by dispersal of the gas disc, since the observed planets are all
expected to be in the regime where runaway gas accretion can occur (e.g. Pollack et al 1996;
Papaloizou \& Nelson 2005) -- i.e. in the saturnian or sub-saturnian mass range. 

The models we have presented are in essence the simplest we can compute, so our intention
has not been to provide rigorous fits to the observational data. Instead, our aim was to demonstrate
the plausibility of the core formation, migration and gas accretion picture for each of the
Kepler circumbinary planets. Our simulations demonstrate the importance of the circumbinary
disc structure for determining the final parameters of forming circumbinary planets, and it 
is clear that improved understanding of these systems will require more in-depth modelling
that includes radiation transport, self-gravity and additional processes that determine
the detailed stucture of the circumbinary disc. These issues will be examined in a future
publication.

\begin{acknowledgements}
Computer time for this study was provided by the computing facilities MCIA (M\'esocentre de Calcul Intensif Aquitain) of the Universit\'e de Bordeaux and by HPC resources from GENCI-cines (c2012046957).

\end{acknowledgements}

\end{document}